\documentclass[pra,twocolumn]{revtex4}
\usepackage{tikz}
\usetikzlibrary{arrows}
\usepackage{graphicx}
\usepackage{dcolumn}
\usepackage{bm}
\usepackage{amsmath}
\usepackage{amsfonts}
\usepackage{psfrag}
\usepackage{graphicx}
\usepackage{color}

\def\bra#1{\mathinner{\langle{#1}|}} 
\def\ket#1{\mathinner{|{#1}\rangle}}

\newcommand{\Eq}[1]{Eq. (\ref{#1})}

\begin{document}

\title{Light-matter decoupling in the deep strong coupling regime:\\ The breakdown of the Purcell effect}

\author{Simone \surname{De Liberato}}
\affiliation{School of Physics and Astronomy, University of Southampton, Southampton, SO17 1BJ, United Kingdom}

\begin{abstract} 
Improvements both in the photonic confinement and in the emitter design have led to a steady increase in the strength of the light-matter coupling in cavity quantum electrodynamics experiments.
This has allowed to access interaction-dominated regimes in which the state of the system can only be described in terms of mixed light-matter excitations. 
Here we show that, when the coupling between light and matter becomes strong enough, this picture breaks down, and light and matter degrees of freedom totally decouple.  A striking consequence of such a counter-intuitive phenomenon is that the Purcell effect is reversed and the spontaneous emission rate, usually thought to increase with the light-matter coupling strength, plummets instead for large enough couplings. 
\end{abstract}

\maketitle

\section{Introduction}
The Purcell effect allows us to engineer the spontaneous emission rate of an emitter by tailoring its photonic environment \cite{Purcell46}. In particular, using resonant photonic cavities with  narrow densities of states, it is possible to greatly enhance the efficiency of photonic devices. To this aim, many  varieties of cavities have been perfected, eventually giving rise to the research field known as  cavity quantum electrodynamics (CQED) \cite{Haroche}. 

One of the key parameters to characterise a CQED setup is the strength of the coupling between light and matter, quantified by the vacuum Rabi frequency $\Omega$. In the weak coupling regime the photonic confinement only amounts to a modification of the spontaneous emission rate and the effect of the light-matter coupling can be described in terms of transitions between uncoupled light and matter states. This ceases to be true when $\Omega$ becomes larger than the linewidths of the light and matter excitations. 
The system is then said to be in the strong coupling regime and, while the spontaneous emission rate still follows a Purcell-like dependency proportional to $\Omega^2$,  we can correctly describe the system only in terms of the dressed eigenstates of the full light-matter Hamiltonian \cite{Kavokin}. 
If the coupling is increased even further, $\Omega$ eventually becomes a non-negligible fraction of the bare frequency of the electronic transition that couples to light, $\omega_0$. This marks the onset of a different regime, called the ultrastrong coupling (USC) regime \cite{Ciuti05,Devoret07}. As the normalised coupling $\frac{\Omega}{\omega_0}$ is the relevant small parameter in the perturbative expansion of the light-matter interaction, in such a regime non-perturbative effects start  to appear and the spontaneous emission rate eventually saturates \cite{Ciuti06}.

When Ciuti, Carusotto, and Bastard published the first theoretical description of the USC regime \cite{Ciuti05}, they limited themselves to values of the normalised coupling smaller than $1$.
This choice was more than reasonable at a time when the largest observed value was of the order of $\frac{\Omega}{\omega_0}\simeq 0.02$ \cite{Dini03}. Still, the fascinating new physics observable in the USC regime, ranging from the dynamical Casimir effect \cite{DeLiberato07,Auer12}, to superradiant phase transitions \cite{Lambert04,Nataf10b}, and ultra-efficient light emission \cite{Ciuti06,DeLiberato08}, stimulated  considerable experimental efforts \cite{Anappara09,Todorov10,Niemczyk10,Muravev11,Schwartz11,Geiser12} that have led to the observation of ever increasing values of the normalised coupling in many different systems, with a present world record $\frac{\Omega}{\omega_0}=0.58$ \cite{Scalari12}.
As a consequence, new theoretical investigations, taking a further leap forward, are starting to study what happens when  the coupling increases even further and $\frac{\Omega}{\omega_0} > 1$ \cite{Casanova10,Hagenmuller10,Beaudoin11,Bamba13}. Judging from recent experimental improvements this regime, usually referred to as deep strong coupling (DSC), will soon be experimentally accessible.

In this paper we will prove that in the DSC regime light and matter effectively decouple: the larger $\Omega$ becomes, the smaller the effective coupling is, such that for large enough couplings no energy is exchanged between light and matter degrees of freedom. One of the most striking consequences of this result is that 
the spontaneous emission rate, thought until now to monotonically increase with the strength of the light-matter coupling, dramatically decreases for suitably large values of $\Omega$.  

In Sec. \ref{LMDec} we will prove that the light-matter decoupling effect is a general consequence of the form of the light-matter coupling Hamiltonian in CQED . In Sec. \ref{SpecSys} we will give a quantitative example of such a decoupling effect by studying a specific CQED model and showing, in Sec. \ref{PurcellBreak}, how the light-matter decoupling leads to a breakdown of the Purcell effect.

\begin{figure}[t]
\begin{center}
\includegraphics[width=9.0cm]{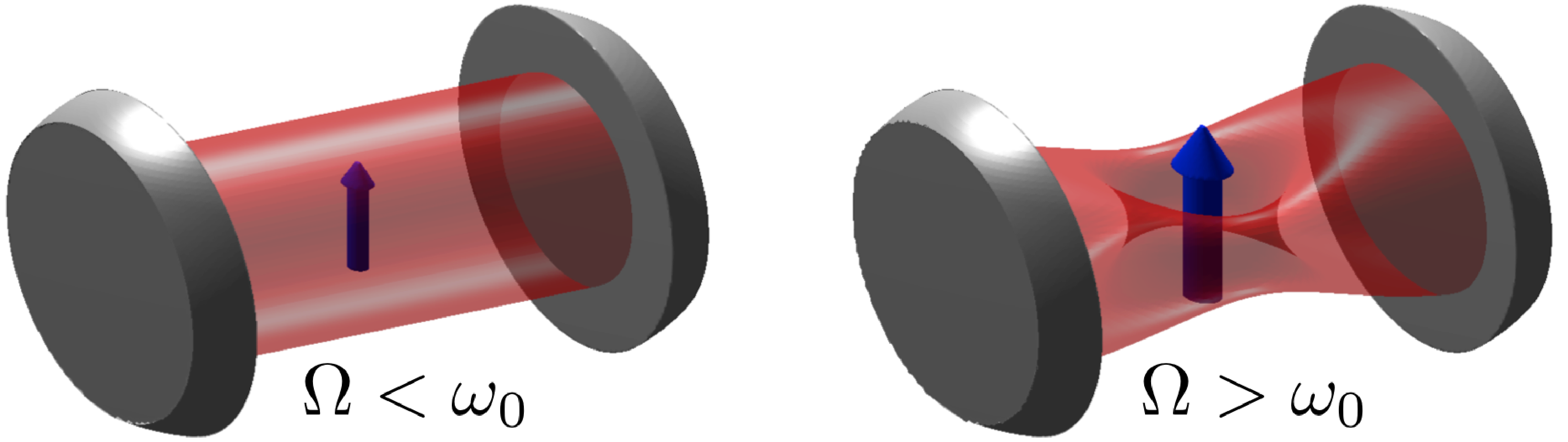}
\caption{\label{Sketch1} A pictorial representation of the light matter decoupling is shown. When the dipole increases, the electric field vanishes at its location.}
\end{center}
\end{figure}

\begin{figure}[t]
\begin{center}
\includegraphics[width=9.0cm]{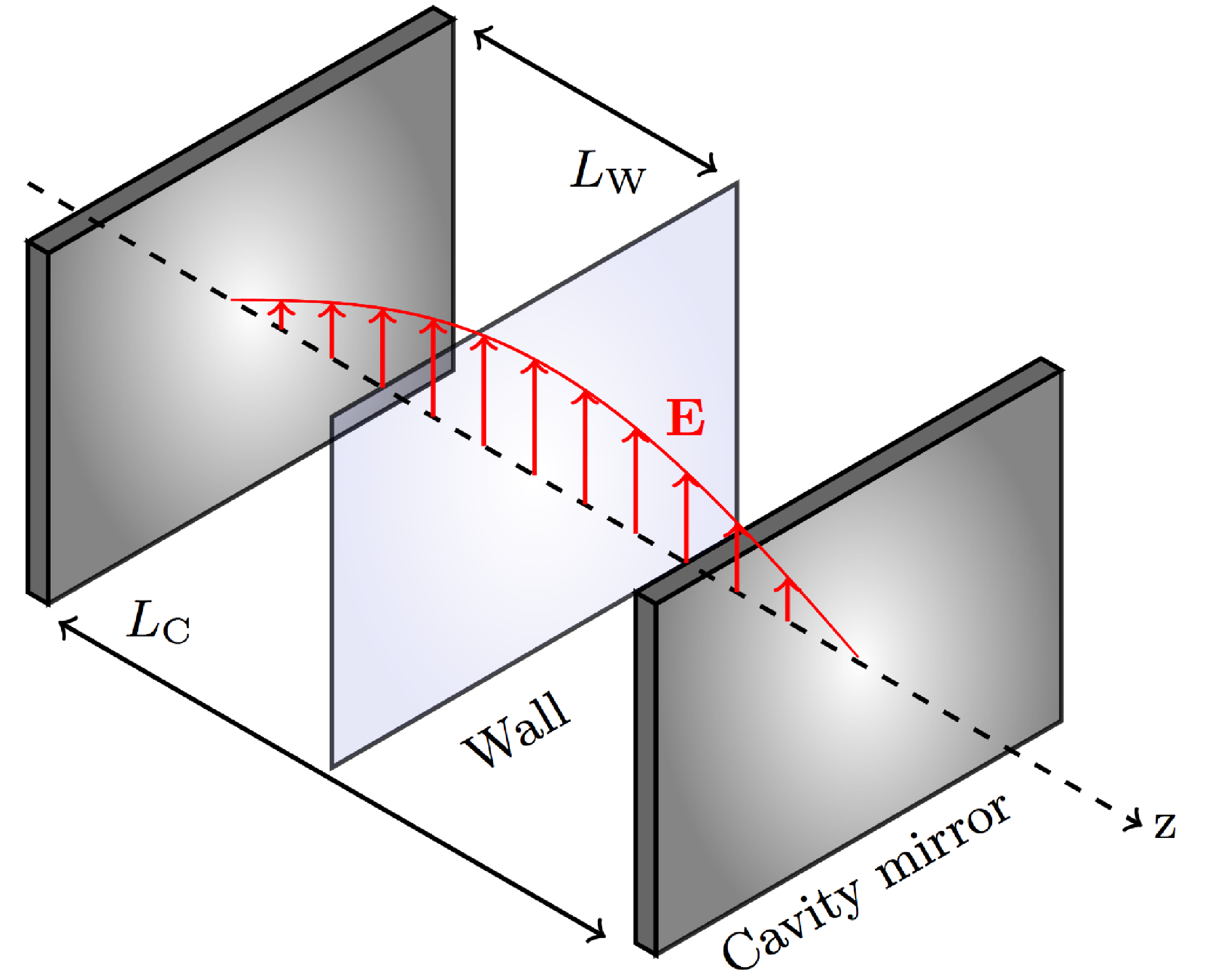}
\caption{\label{Sketch2} A sketch of the specific system studied. A wall of dipoles is enclosed into a planar metallic cavity of length $L_{\text{C}}$, at a distance $L_{\text{W}}$ from one of the mirrors.}
\end{center}
\end{figure}

\section{Light-matter decoupling}
\label{LMDec}
The light-matter decoupling in the DSC regime is due to the general form of the Hamiltonian describing a set of electrons coupled to the electric field of a cavity
\begin{eqnarray}
\label{H1q}
H_{\text{LM}}&=&H_0+ \sum_{j} \frac{e}{m}\mathbf{p}_j \cdot \mathbf{A}(\mathbf{r}_j)+ \frac{e^2}{2m}\mathbf{A}(\mathbf{r}_j)^2,
\end{eqnarray}
where $H_0$ is the free Hamiltonian of electrons and cavity field, $\mathbf{r}_j$ and $\mathbf{p}_j$ are position and momentum of electron $j$, and  $\mathbf{A}(\mathbf{r})$ is the vector potential in $\mathbf{r}$. If $H_0$ gives rise to a dipolar allowed electronic transition of frequency $\omega_0$, whose coupling to the cavity field, due to the $\mathbf{p} \cdot \mathbf{A}(\mathbf{r})$ term, is of order $\Omega$, by the TRK sum rule \cite{Nataf10b} the coefficient of the $\mathbf{A}(\mathbf{r})^2$ term will be at least of order $\frac{\Omega^2}{\omega_0}$. This means that in the DSC regime $\frac{\Omega}{\omega_0}>1$, eventually the energy contribution of this last quadratic term, always positive, becomes the dominant one. The low energy part of the spectrum will thus be composed of modes that minimise the $\mathbf{A}(\mathbf{r})^2$ term: either  
pure matter excitations, with a vanishing photonic component, or excitations whose electric field configuration presents nodes  at the locations of the dipoles.
 The dipoles effectively expel the field in the DSC regime and thus energy exchanges between light and matter degrees of freedom, that are always due to local interactions, vanish. A pictorial representation of this phenomenon is shown in Fig. (\ref{Sketch1}). 
It is worthwhile to notice that the light-matter decoupling depends on the presence of the $\mathbf{A}(\mathbf{r})^2$ term, and it could thus be modified if such a term is somehow altered. A more in-depth discussion of this point can be found in Appendix \ref{AboutA2}.

\section{Application to a specific system}
\label{SpecSys}
In the following, in order to give a quantitative example of this striking decoupling effect and of its consequences, we will  study in detail the specific CQED system sketched in Fig. (\ref{Sketch2}), that is a metallic planar cavity of length  $L_{\text{C}}$ enclosing a two dimensional wall of dipoles at distance $L_{\text{W}}$ from one of the mirrors. We will consider only normal incidence and, in order to be able to study the coupling in real space, we will  take into account all the excited modes of the cavity.
The choice of such a system is motivated by the fact that, in its generality, it is a reasonably good toy model for almost all the experiments in which the USC regime has been observed to date \cite{Anappara09, Todorov10, Schwartz11,Muravev11,Geiser12,Scalari12}, the only exception are superconducting circuits \cite{Niemczyk10}, in which a single dipole couples to the cavity field.

\psfrag{T1}[T][B][0.8]{\hspace{-2.0cm}$\mathbf{(a)}$}
\psfrag{T2}[T][B][0.8]{\hspace{-1.5cm}$\mathbf{(b)}$}
\psfrag{T3}[T][B][0.8]{\hspace{-1.5cm}$\mathbf{(c)}$}
\psfrag{T4}[T][B][0.8]{\hspace{-1.5cm}$\mathbf{(d)}$}
\psfrag{T5}[T][B][0.8]{\hspace{-2.0cm}$\mathbf{(e)}$}
\psfrag{T6}[T][B][0.8]{\hspace{-1.5cm}$\mathbf{(f)}$}
\psfrag{T7}[T][B][0.8]{\hspace{-1.5cm}$\mathbf{(g)}$}
\psfrag{T8}[T][B][0.8]{\hspace{-1.5cm}$\mathbf{(h)}$}

\psfrag{X1}[t][c][0.8]{$\Omega/\omega_0$}
\psfrag{X5}[t][c][0.8]{$\Omega/\omega_0$}
\psfrag{X2}[t][c][0.8]{$z/L_{\text{C}}$}
\psfrag{X3}[t][c][0.8]{$z/L_{\text{C}}$}
\psfrag{X4}[t][c][0.8]{$z/L_{\text{C}}$}
\psfrag{X6}[t][c][0.8]{$z/L_{\text{C}}$}
\psfrag{X7}[t][c][0.8]{$z/L_{\text{C}}$}
\psfrag{X8}[t][c][0.8]{$z/L_{\text{C}}$}
\psfrag{Y1}[B][t][0.8]{$\omega/\omega_{\text{c}}$}
\psfrag{Y5}[B][t][0.8]{$\omega/\omega_{\text{c}}$}
\psfrag{Y2}[B][t][0.8]{$\lvert \mathbf{E}(z)\lvert$ (Arb. units)}
\psfrag{Y6}[B][t][0.8]{$\lvert \mathbf{E}(z)\lvert$ (Arb. units)}
\psfrag{Y3}[B][c][0.8]{}
\psfrag{Y4}[B][c][0.8]{}
\psfrag{Y7}[B][c][0.8]{}
\psfrag{Y8}[B][c][0.8]{}

\begin{figure*}[t]
\begin{center}
\hspace{-1.0cm}
\includegraphics[width=5.0cm]{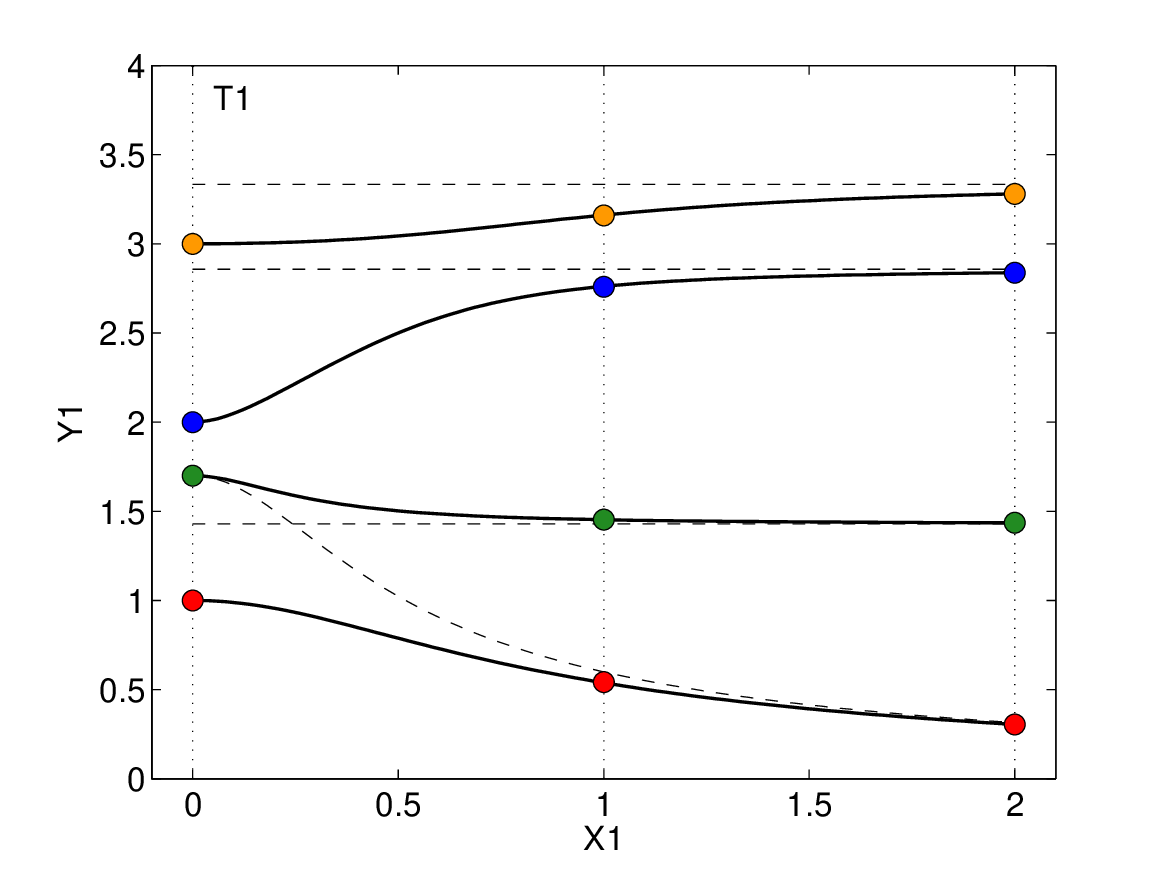}
\hspace{-0.45cm}
\includegraphics[width=5.0cm]{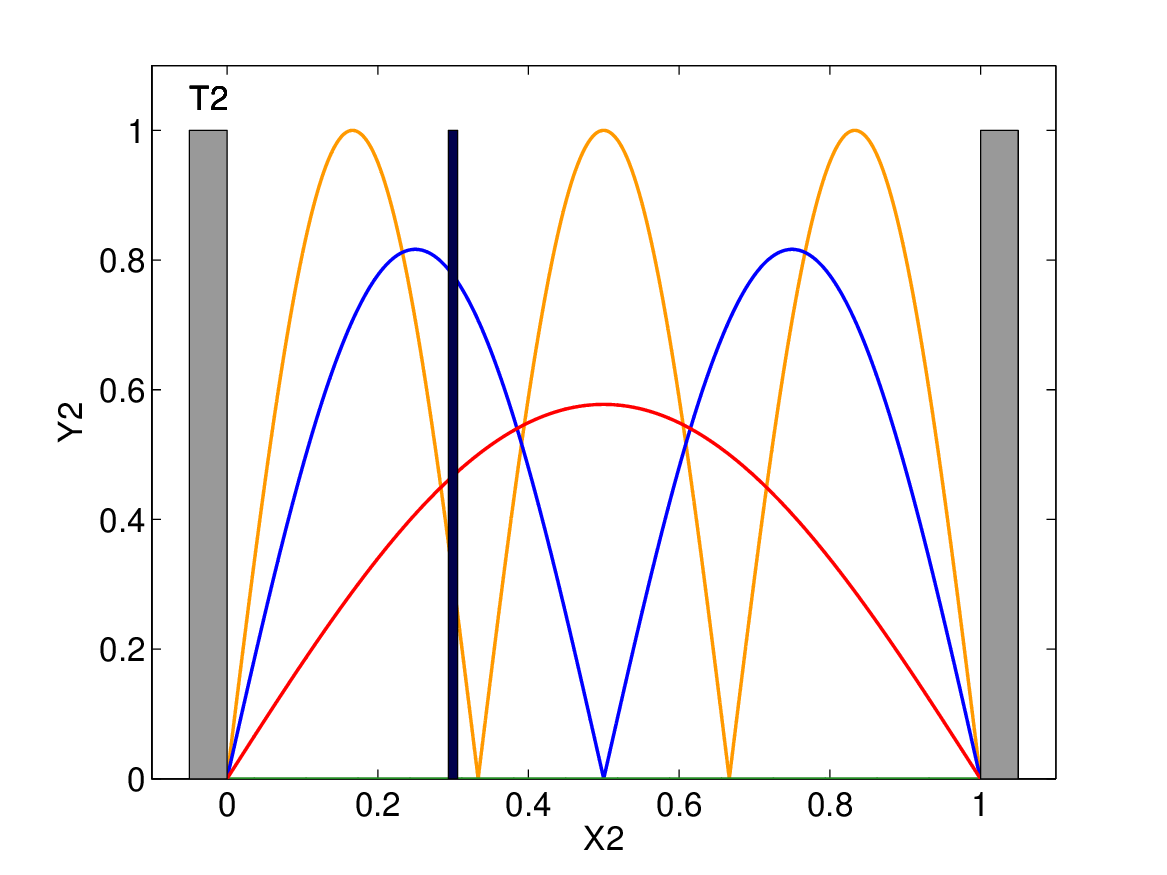}
\hspace{-0.65cm}
\includegraphics[width=5.0cm]{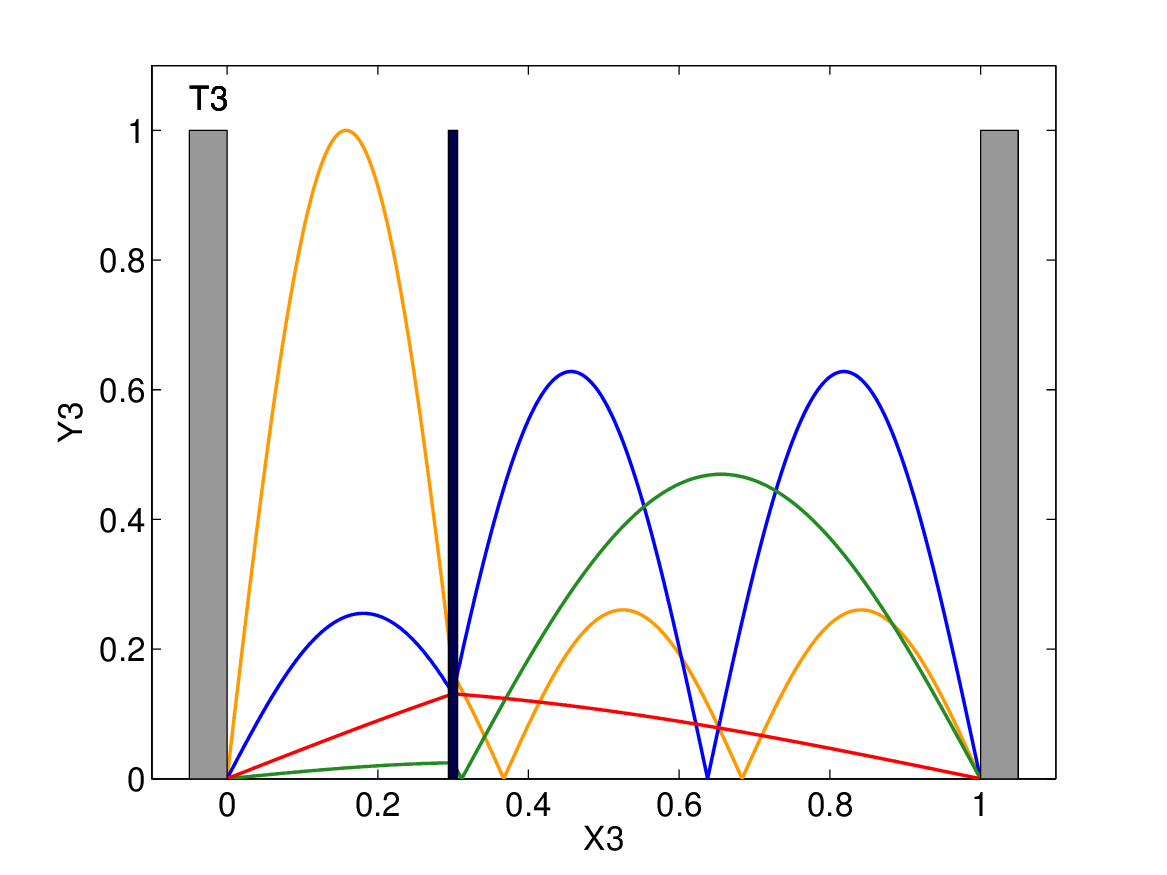}
\hspace{-0.65cm}
\includegraphics[width=5.0cm]{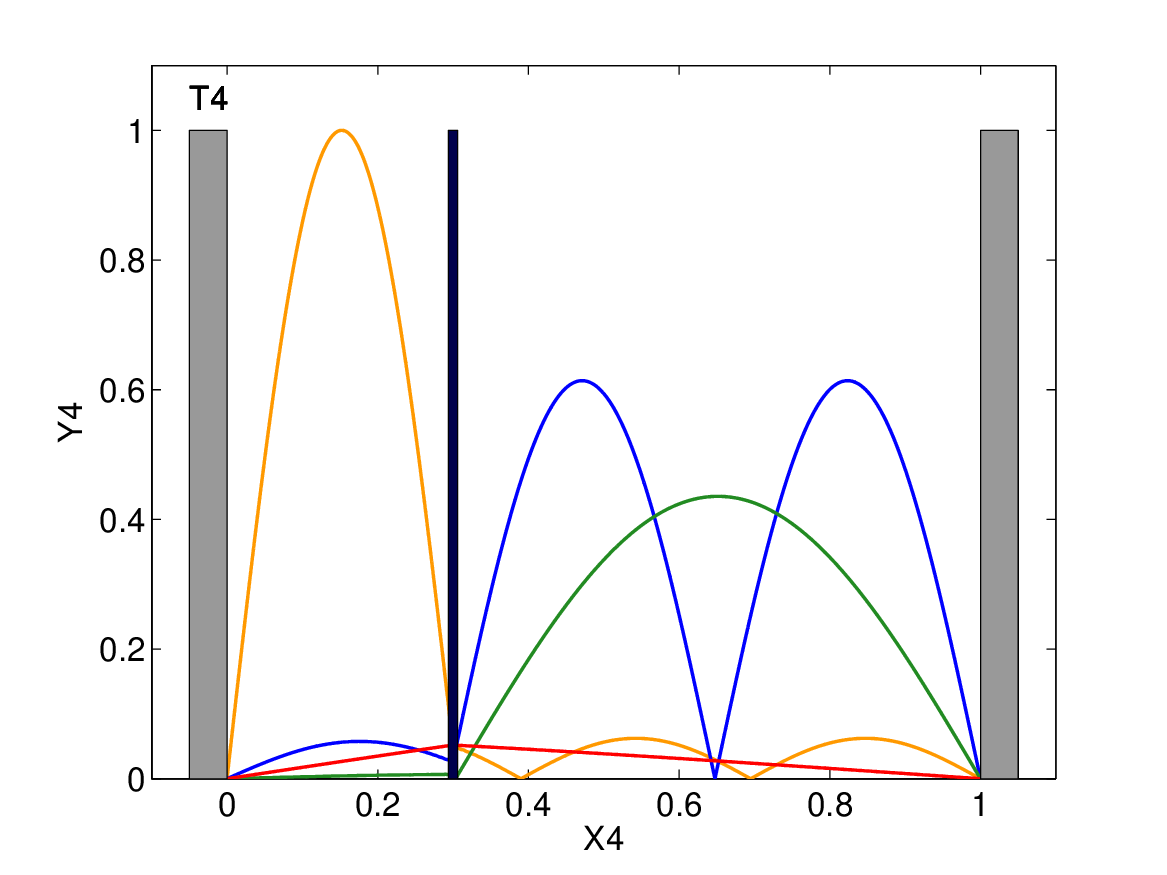}
\end{center}

\begin{center}
\hspace{-1.0cm}
\includegraphics[width=5.0cm]{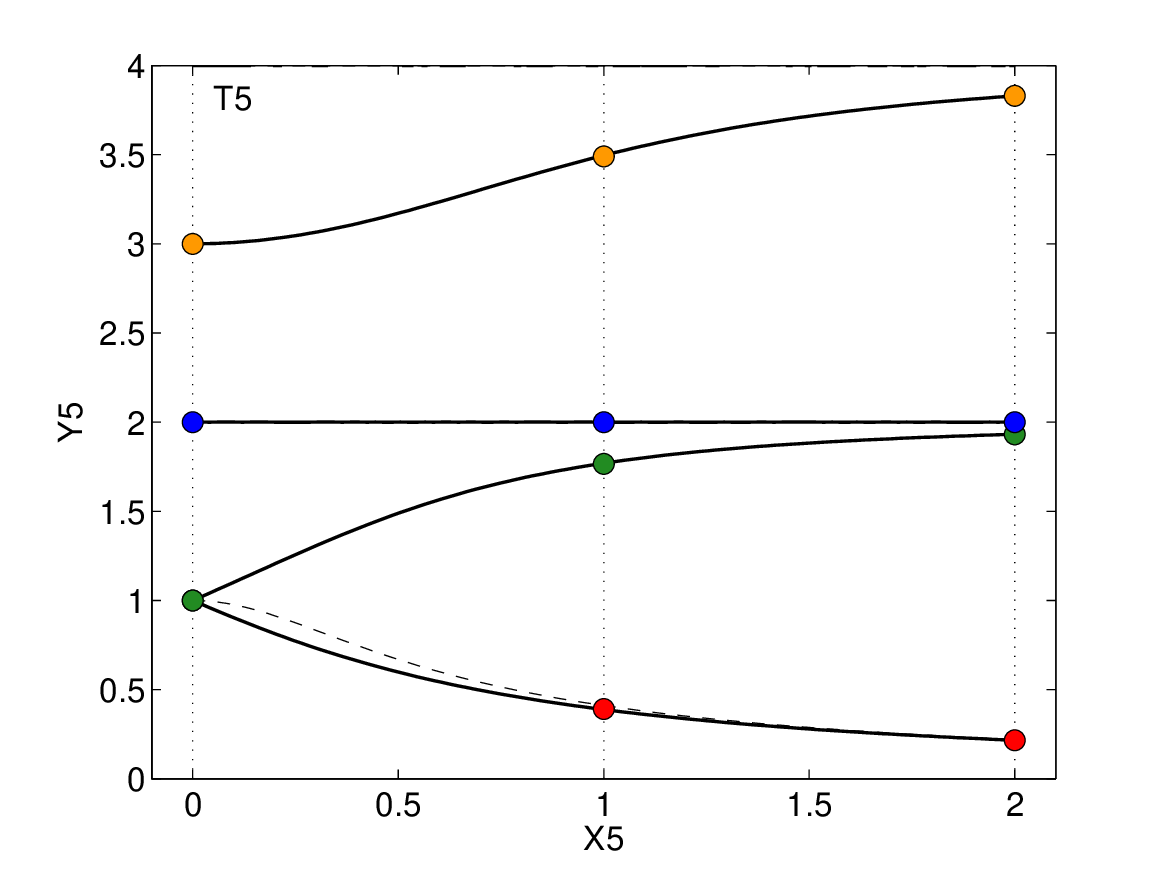}
\hspace{-0.45cm}
\includegraphics[width=5.0cm]{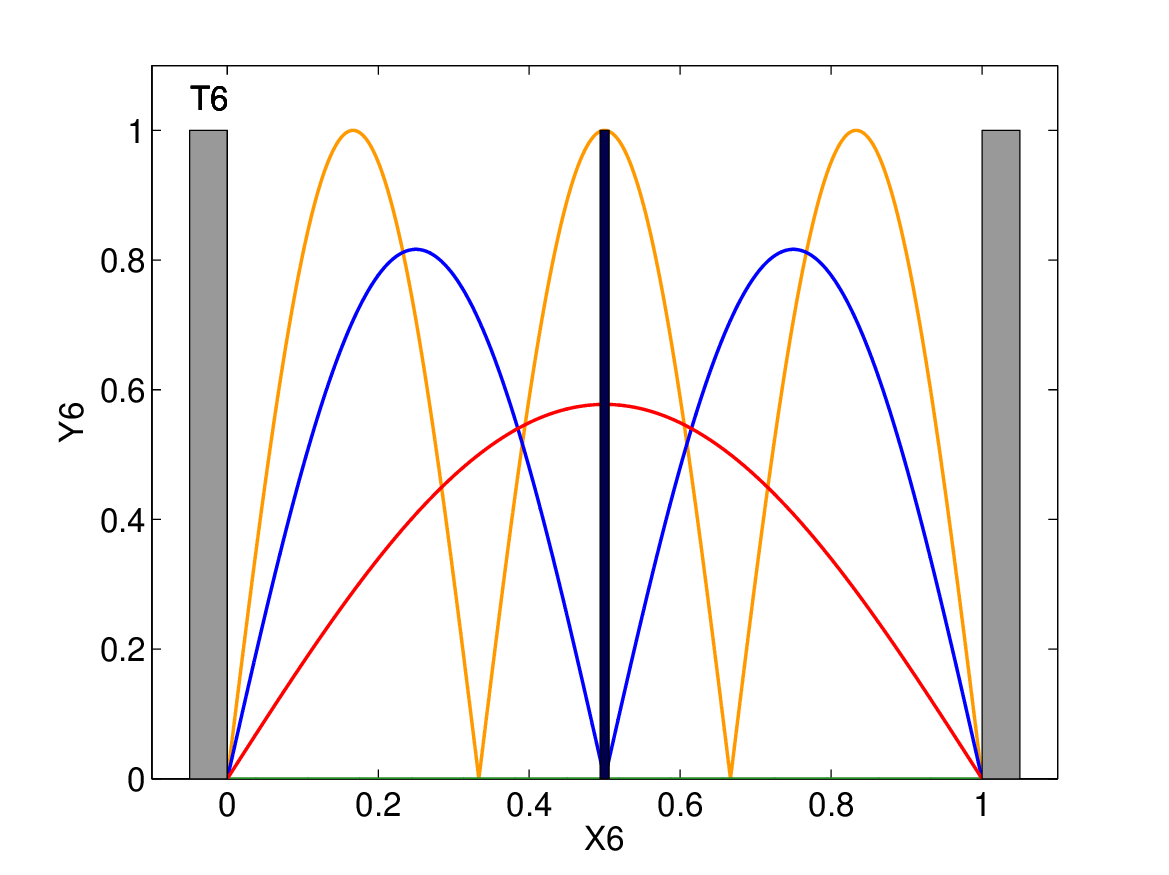}
\hspace{-0.65cm}
\includegraphics[width=5.0cm]{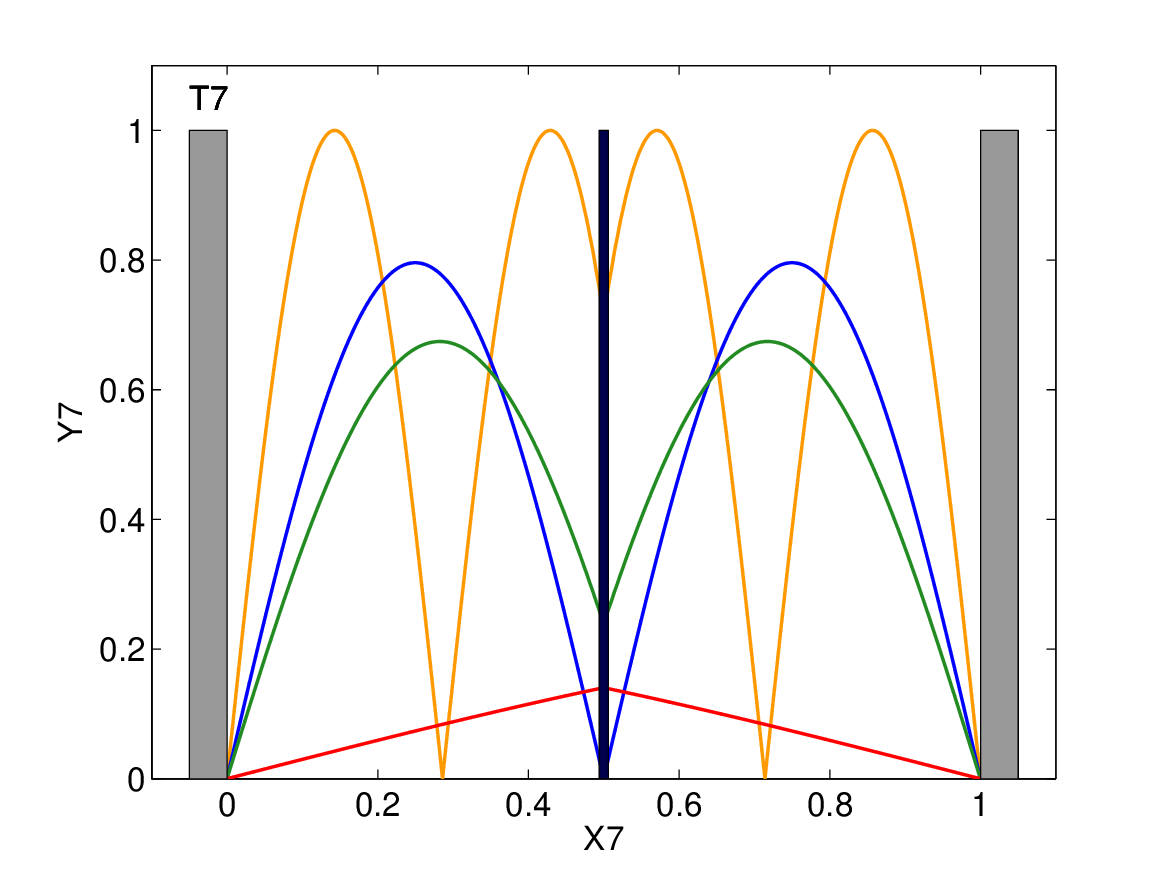}
\hspace{-0.65cm}
\includegraphics[width=5.0cm]{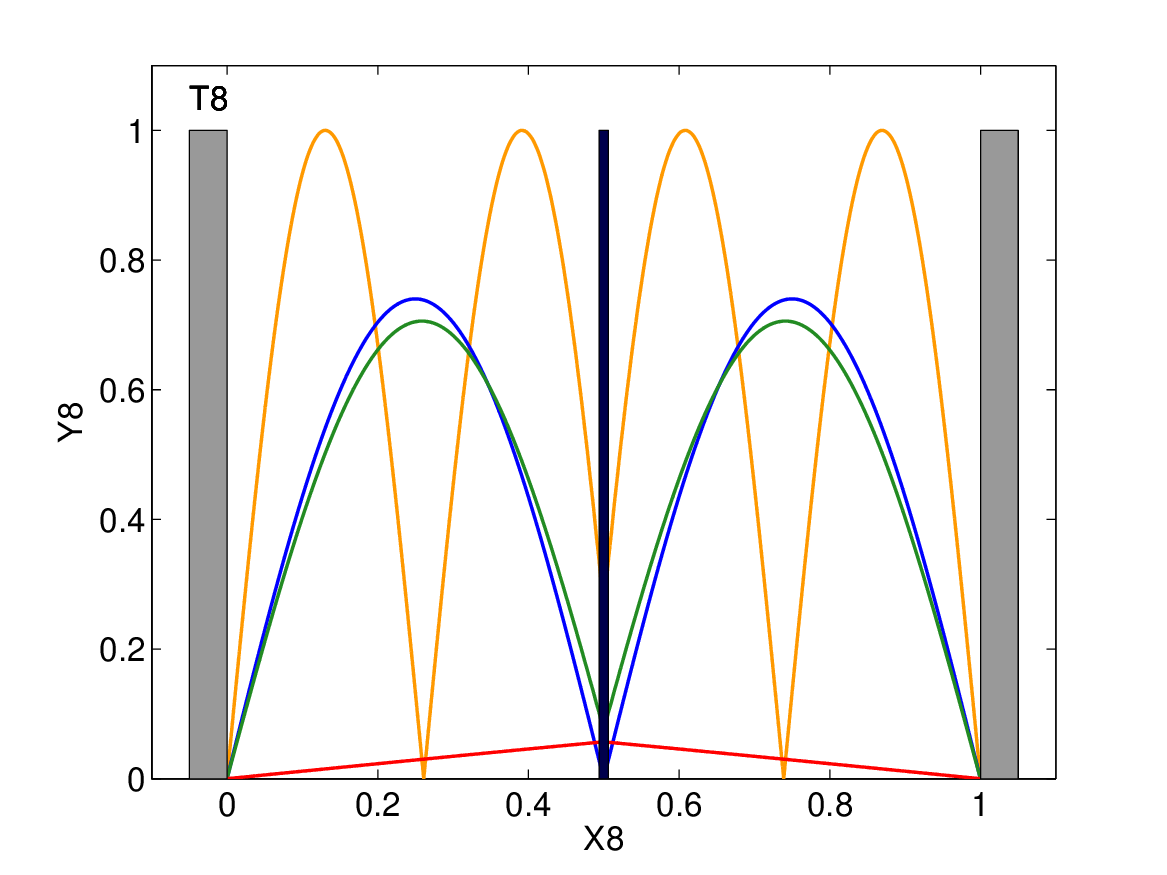}
\caption{\label{MainFig} 
We plot the frequency of the first few modes as a function of the normalised coupling  (a,e) and the profile of the electric field for $\frac{\Omega}{\omega_0}=0$ (b,f),
$\frac{\Omega}{\omega_0}=1$ (c,g), and $\frac{\Omega}{\omega_0}=2$ (d,h).
Images of the first row (a-d) have parameters $\omega_0=1.7\omega_{\text{c}}$ and $l=0.3$, while for the ones in the second row (e-h) $\omega_0=\omega_{\text{c}}$ and $l=0.5$.
In (a,e) the dashed lines correspond to the asymptotic values in \Eq{Eigs}, that is a mode going toward zero frequency for high couplings and the modes obtained when substituting the dipole wall with a metallic mirror. In the figures showing the profile of the electric field, a graphical representation of the mirrors (two grey shaded areas) and of the dipole wall (a thick vertical line) are added for clarity.
}
\end{center}
\end{figure*}

Supposing that only a single electronic transition of frequency $\omega_0$ couples to light, such that the TRK sum rule is saturated, we can write the light matter Hamiltonian as 
\begin{eqnarray}
\hspace{-0.3cm}
\label{Hfull}
H_{\text{LM}}&=&\hbar\omega_0 b^{\dagger}b+ \sum_n  \hbar n \omega_{\text{c}} a_n^{\dagger}a_n +  \sum_{n} \hbar  \Omega_n
(b^{\dagger}+b)(a^{\dagger}_n+a_n)\nonumber\\&&+ \sum_{n,m} \frac{\hbar\Omega_n\Omega_m}{\omega_0}(a^{\dagger}_n+a_n)(a^{\dagger}_m+a_m),
\end{eqnarray}
where $b$ and $a_n$ are respectively the annihilation operator for an electronic excitation and for the  $n^{\text{th}}$ photonic mode of the cavity,
whose coupling coefficient is $\Omega_n=\frac{\Omega \sin(\pi l n)}{\sqrt{n}}$, with $l=\frac{L_{\text{W}}}{L_{\text{C}}}$.
Here and in the following, latin letters run over the modes of the empty cavity, while greek ones over the modes of the coupled system.
Given the infinite in-plane extension of the system, the matter operator will satisfy bosonic commutation relations $\lbrack b,\,b^{\dagger}\rbrack=1$  \cite{Ciuti05}.
As shown in the Appendix \ref{Diagonalisation}, the spectrum of $H_{\text{LM}}$ is given by the solutions of the equation 
\begin{eqnarray}
\label{Eqsp}
\omega_0^2-\omega^2=2\pi \Omega^2 \frac{\omega}{\omega_0} \frac{\sin(\pi l \frac{\omega}{\omega_{\text{c}}})\sin(\pi (1-l) \frac{\omega}{\omega_{\text{c}}})}{\sin(\pi \frac{\omega}{\omega_{\text{c}}})},
\end{eqnarray}
that can be calculated analytically in the asymptotic $\frac{\Omega}{\omega_0}\gg 1$ limit as
\begin{eqnarray}
\label{Eigs}
\omega&=&\frac{\omega_0}{\sqrt{1+\frac{2\pi^2\Omega^2l(1-l)}{\omega_0\omega_{\text{c}}}}}\nonumber\\
\omega&=&\frac{n\omega_{\text{c}}}{l},\quad \omega=\frac{n\omega_{\text{c}}}{1-l},\quad n\in\mathbb{N}.
\end{eqnarray}
The first line of \Eq{Eigs} describes a mode going toward zero frequency, while the second line the modes of two cavities of length $L_{\text{W}}$ and $L_{\text{C}}-L_{\text{W}}$.
In Fig. (\ref{MainFig}) (a,e), we plot both the exact solutions of \Eq{Eqsp} (solid lines) and the asymptotic values from \Eq{Eigs} (dashed lines).
\psfrag{XH}[t][c][1]{$\Omega/\omega_0$}
\psfrag{YH}[b][t][1]{$\chi$}
\begin{figure}[t]
\begin{center}
\includegraphics[width=8.0cm]{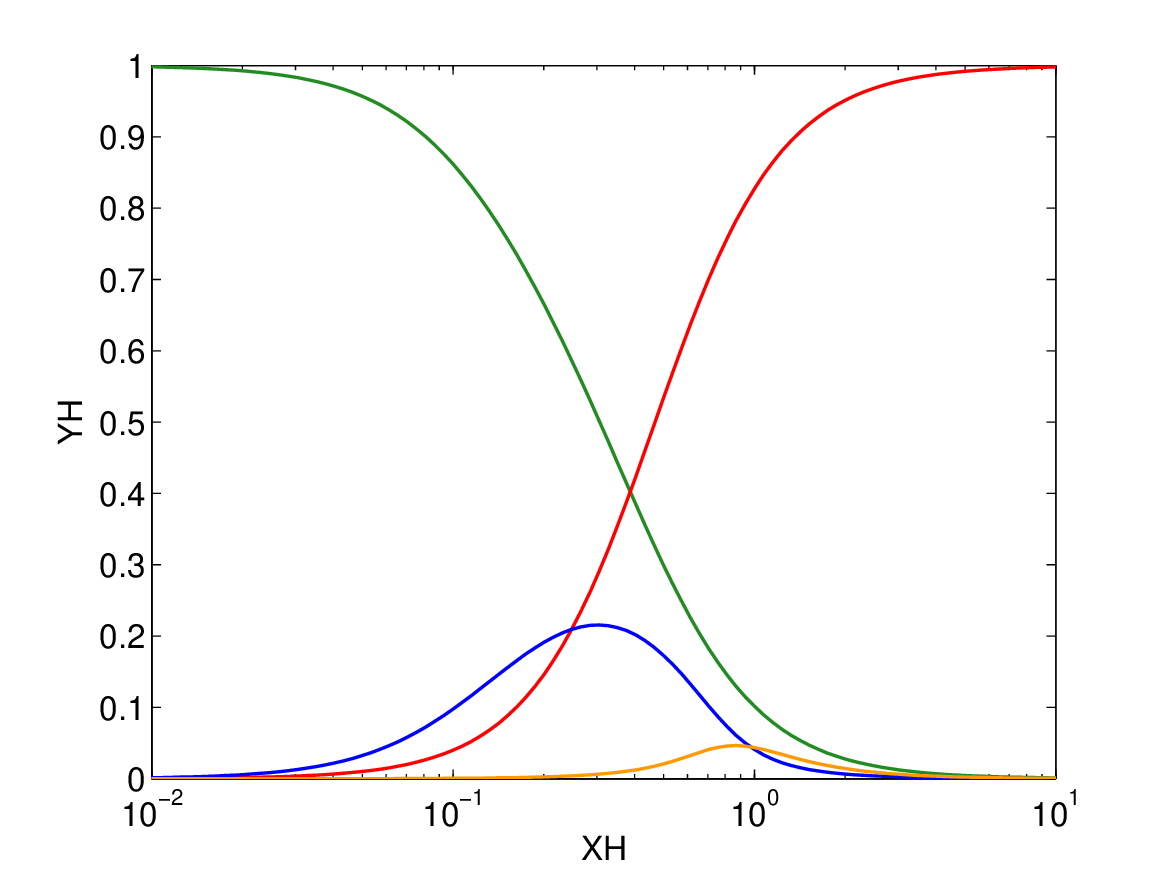}
\end{center}
\caption{\label{HBcoeff} Total matter fraction for the four lowest polaritonic modes. The parameters and the colors correspond to Fig. (\ref{MainFig}) (a). The light-matter decoupling effect in the DSC regime is clearly visible, with the system that reverts to an uncoupled state for $\Omega/\omega_0>1$.} 
\end{figure}
In order to better understand the nature of the modes in \Eq{Eigs} we need to calculate their electric field distribution. To this aim, we perform a Hopfield-Bogoliubov transformation of $H_{\text{LM}}$, detailed in Appendix \ref{HBdiag}, that allows us to write the  Hamiltonian in diagonal form as 
\begin{eqnarray}
\label{Hpol}
H_{\text{LM}}&=&\sum_{\mu} \hbar \omega_{\mu} p_{\mu}^{\dagger}p_{\mu},
\end{eqnarray}
where the $\omega_{\mu}$ are the solutions of \Eq{Eqsp}, $\mu$ runs over all the eigenstates of $H_{\text{LM}}$, and the bosonic polaritonic $p_{\mu}$ operators are linear superpositions of both creation and annihilation operators for the light and matter modes
\begin{eqnarray}
\label{pdec}
p_{\mu}&=&U_{\mu 0}b+\sum_n U_{\mu n}a_n+ V_{\mu 0}b^{\dagger}+\sum_n V_{\mu n}a_n^{\dagger}.
\end{eqnarray}
Inverting \Eq{pdec},  we can write the electric field as a function of the polaritonic $p_{\mu}$ operators
\begin{eqnarray}
\label{E}
\mathbf{E}(z)&=&i\mathbf{e}_{\parallel}\sum_n \sqrt{\frac{\hbar \omega_n}{\epsilon_0 L_{\text{C}} S}}  \sin(\frac{z n \pi}{L_{\text{C}}}) ( a_n^{\dagger}-a_n)\\&=&\nonumber
i\mathbf{e}_{\parallel}\sum_{n,\mu} \sqrt{\frac{\hbar \omega_n}{\epsilon_0 L_{\text{C}} S}}  \sin(\frac{z n \pi}{L_{\text{C}}}) (U_{\mu n}+V_{\mu n}) ( p_{\mu}^{\dagger}-p_{\mu}),
\end{eqnarray}
where $S$ is the quantization surface and $\mathbf{e}_{\parallel}$ a polarization vector orthogonal to the cavity axis.
Calling $\ket{G}$ the ground state of $H_{\text{LM}}$, defined by $p_{\mu}\ket{G}=0$, and $\mathbf{E}^+(z)$ and $\mathbf{E}^-(z)$ the positive and negative frequency components of \Eq{E},  the field profile inside the cavity $\lvert \mathbf{E}_{\mu}(z)\lvert$ for the polaritonic mode $\mu$ is given by the photodetection signal
\begin{eqnarray}
\label{E2}
\lvert \mathbf{E}_{\mu}(z) \lvert^2&=&\bra{G}p_{\mu} \mathbf{E}^+(z)\mathbf{E}^-(z)p^{\dagger}_{\mu}\ket{G}.
\end{eqnarray}
In Fig. (\ref{MainFig}) we plot the field profile for the first few modes and for different values of $\Omega$ and $\omega_0$. 
Our results confirm the predictions from the general argument stated in the first part of the present work: the electric field of all the modes except the lowest lying one vanishes at the locations of the dipoles, while the lowest mode field  has a maximum at the dipole wall, but its overall intensity vanishes. 
In the limit of infinite coupling, the polaritonic modes become either static polarizations confined in the dipole wall, or the modes  of two empty cavities of length $L_{\text{W}}$ and $L_{\text{C}}-L_{\text{W}}$. These are the modes we would obtain imposing metallic boundary conditions at the location of the wall: as the electric field vanishes on the dipoles, a dipole wall effectively behaves as a metallic mirror. Notice that, as clearly shown in Fig. (\ref{MainFig}), relatively small values of the normalised coupling suffice to access these asymptotic behaviours.

It is interesting to analyse the relative weights of the light and matter coefficients of the polaritonic operators in \Eq{pdec}. In Fig. (\ref{HBcoeff}) we plot the normalised total matter component 
\begin{eqnarray}
\label{chiM}
\chi_{\mu}&=&\frac{U_{\mu 0}^2+V_{\mu 0}^2}{U_{\mu 0}^2+V_{\mu 0}^2+\sum_n (U_{\mu n}^2+V_{\mu n}^2)},
\end{eqnarray}
for the modes $\mu$ whose coupling dependency is shown in Fig. (\ref{MainFig}) (a). While this quantity does not have an operational physical meaning, due to the fact that the polaritonic operators are $\eta$-normalised (see Appendix \ref{HBdiag} for details), it does converge to 0 or 1 for, and only for, pure radiation and pure matter modes respectively.
Figure (\ref{HBcoeff}) highlights the symmetry between the weak coupling and the DSC regimes. Light and matter modes, although shifted, become completely decoupled for large enough couplings.
As a further proof of the solidity of the presented theory, in the Appendix \ref{ClassicalTheory}  we show how its  main components,  including the polaritonic eigenfrequencies in \Eq{Eqsp} and the profile of the modes electric field in Fig. (\ref{MainFig}), can be obtained using a completely classical transfer matrix approach. 

\section{The breakdown of the Purcell effect}
\label{PurcellBreak}
This effective light-matter decoupling strongly influences the luminescence of the system. Electric pumping can only excite the matter component of a polariton, while the probability of it decaying by emitting a photon out of the cavity is a function of
its electromagnetic component. As the modes of the coupled system, for large enough couplings, become either matter or radiation, we expect that, in stark contrast to what reported in the literature until now \cite{Ciuti06,DeLiberato08,Jouy10},  the spontaneous emission rate, and thus the electroluminescence efficiency, should decrease, not increase, when the light-matter coupling increases beyond a certain threshold.

\psfrag{XEff}[t][c][1]{$\Omega/\omega_0$}
\psfrag{YEff}[b][t][1]{$\gamma/\omega_0$}

\begin{figure}[t]
\begin{center}
\includegraphics[width=8.0cm]{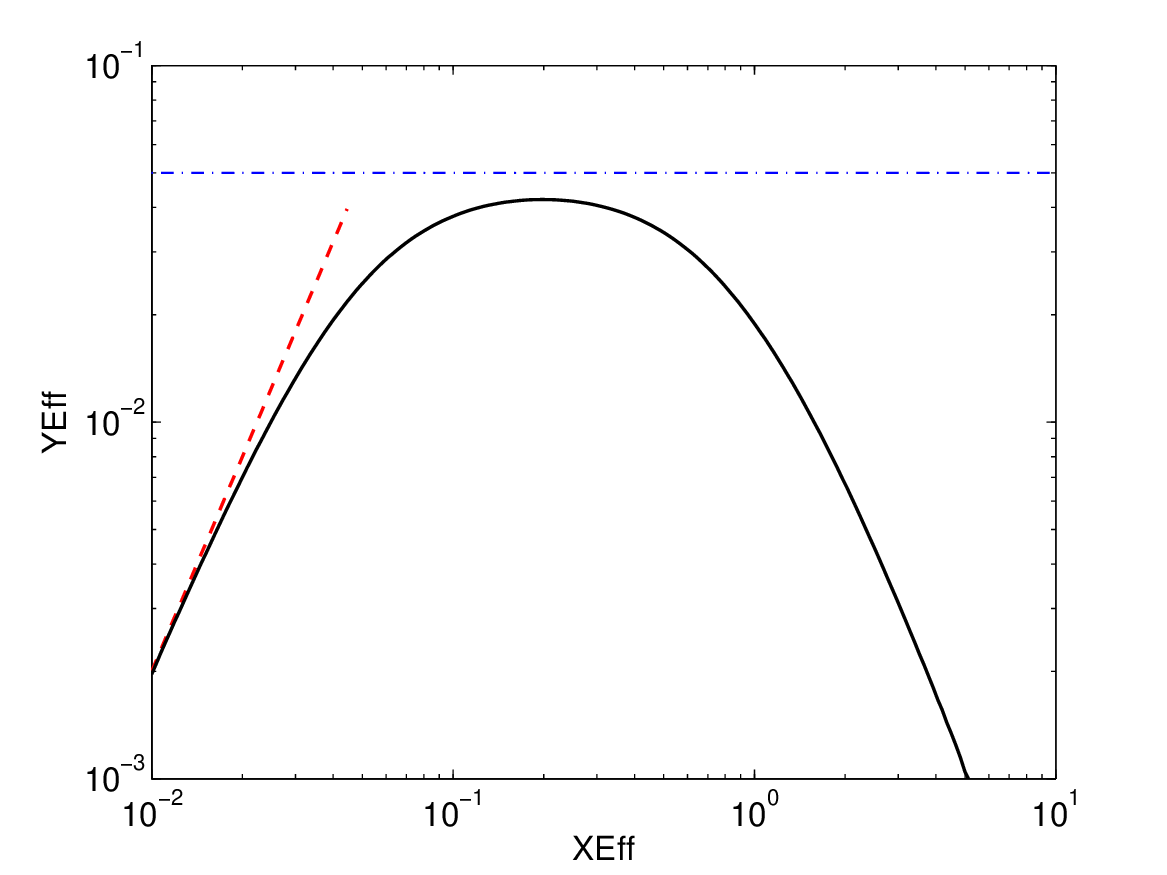}
\end{center}
\caption{\label{FigEff} Spontaneous emission rate as a function of the normalised light-matter coupling. The black solid line is calculated using \Eq{Lum}, the red dashed one is the weak coupling approximation from \Eq{LumP} and the blue dash-dotted one is the saturation value from \Eq{LumS}.  Parameters are $\omega_0=\omega_{\text{c}}$, $l=0.5$, and $\Gamma_{\text{el}}=\Gamma_{\text{ph}}=0.05\omega_0$.}
\end{figure}

In order to prove this point we can generalise the input-output theory developed in Ref. \cite{Ciuti06}, to the case of multiple photonic modes, coupling the intra-cavity modes to two baths of photonic and electronic extra-cavity excitations through the Hamiltonians
\begin{eqnarray}
\label{BPH}
H_{\text{ph}}&=&\hbar\int_0^{\infty} d\omega\, \omega\, \alpha_{\omega}^{\dagger}\alpha_{\omega} +\sum_{n}\kappa_{\text{ph}} \lbrack a_{n}^{\dagger}\alpha_{\omega} +\alpha_{\omega}^{\dagger}a_{n}\rbrack, \nonumber \\
H_{\text{el}}&=&\hbar\int_0^{\infty}  d\omega\, \omega\, \beta_{\omega}^{\dagger}\beta_{\omega} +\kappa_{\text{el}} \lbrack b_{n}^{\dagger}\beta_{\omega} +\beta_{\omega}^{\dagger}b_{n}\rbrack,
\end{eqnarray}
where we considered the coupling coefficients to be frequency independent, as this has recently been shown to give results consistent with a fully microscopic approach derived using Maxwell boundary conditions \cite{Bamba13,DeLiberato13b}.
The presence of extra-cavity modes gives a finite lifetime to intra-cavity light and matter modes, quantified by the loss coefficients $\Gamma_{\text{ph}}=\pi \kappa_{\text{ph}}^2$ and $\Gamma_{\text{el}}=\pi \kappa_{\text{el}}^2$. 
As detailed in Appendix \ref{In-out}, using the standard definitions
\begin{eqnarray}
\label{Bdef}
\alpha_{\omega}^{\text{in}}&\equiv& \lim_{t \rightarrow -\infty}\alpha_{\omega}(t)e^{-i\omega t},\quad \beta_{\omega}^{\text{in}}\equiv \lim_{t \rightarrow -\infty}\beta_{\omega}(t)e^{-i\omega t},\nonumber\\
\alpha_{\omega}^{\text{out}}&\equiv& \lim_{t \rightarrow \infty}\alpha_{\omega}(t)e^{-i\omega t},\quad \beta_{\omega}^{\text{out}}\equiv \lim_{t \rightarrow \infty}\beta_{\omega}(t)e^{-i\omega t}, \quad\quad
\end{eqnarray}
for the input and output fields, one obtains a linear set of equations describing the scattering properties of the system 
\begin{eqnarray}
\left[
\begin{array}{c}
 \alpha^{\text{out}}_{\omega}    \\
 \beta^{\text{out}}_{\omega}
\end{array}
\right]
&=&
\left[
\begin{array}{cc}
 \mathcal{U}^{11}(\omega) &   \mathcal{U}^{12}(\omega)  \\
 \mathcal{U}^{21}(\omega) & \mathcal{U}^{22}(\omega)      
\end{array}
\right]
\left[
\begin{array}{c}
 \alpha^{\text{in}}_{\omega}    \\
 \beta^{\text{in}}_{\omega}
\end{array}
\right].
\end{eqnarray}
Considering an incoherent and frequency-independent electronic input
$\langle  \beta^{\text{in}\dagger}_{\omega}\beta^{\text{in}}_{\omega'} \rangle=\delta(\omega-\omega')I,
$ we can calculate the electroluminescence as 
\begin{eqnarray}
\label{Lum}
\mathfrak{L}&=&\int \lvert \mathcal{U}^{12}(\omega)\lvert^2  \frac{d\omega}{2\pi} I=\gamma I,
\end{eqnarray}
where $\gamma$ is the spontaneous emission rate. A plot of such a quantity can be found in Fig. (\ref{FigEff}), where it is possible to verify that, as predicted in Ref. \cite{Ciuti06},
in the weak coupling regime $\gamma$ is well approximated by the Purcell-like dependency
\begin{eqnarray}
\label{LumP}
\gamma&\simeq&\frac{2\Omega^2}{\Gamma_{\text{el}}+\Gamma_{\text{ph}}},
\end{eqnarray}
while it saturates around the value
\begin{eqnarray}
\label{LumS}
\gamma&\simeq&\frac{2\Gamma_{\text{el}}\Gamma_{\text{ph}}}{\Gamma_{\text{el}}+\Gamma_{\text{ph}}},
\end{eqnarray}
for $\frac{\Omega}{\omega_0}\simeq 0.1$.
The novelty of Fig. (\ref{FigEff}), and one of the main results of the present work, is that, due to light-matter decoupling, the spontaneous emission rate then decreases dramatically.
It is interesting to notice that in Ref. \cite{Ciuti06} a small decrease is indeed observed, at the very upper bound of the considered parameter region, but there such a feature was attributed to a numerical artifact due to the choice of a discontinuous damping. As explained in Appendix \ref{In-out}, we verified numerically that the light-matter decoupling is not qualitatively modified if a continuous damping is used. 

\section{Conclusions}
\label{Conclusions}
In conclusion, we proved that there are intrinsic limits to the effective strength of the light-matter coupling, as in the DSC regime light and matter effectively decouple.
This phenomenon has consequences on the possibility to realise ultraefficient and ultrastrongly coupled light emitters, and it could be exploited to dynamically control the electromagnetic field distribution in optoelectronic devices, by modulating the intensity of the light matter interaction by optical or electrical means \cite{Anappara05,Schwartz11,Porer12}. 

\section{Acknowledgments}
The author acknowledges support of the European Commission under the Marie Curie IEF Program, project BiGExPo.
\appendix

\section{About the $\mathbf{A}(\mathbf{r})^2$ term}
\label{AboutA2}
The discussion presented in the first part of the present work proves that the  light-matter decoupling effect  is due to the presence of the $\mathbf{A}(\mathbf{r})^2$ term in the  Hamiltonian. The fact that such a term hampers the observation of exotic physical phenomena is well known in the community studying the superradiant phase transition, as various no-go theorems prove that Hamiltonians of the same form of the one in \Eq{H1q} do not present a phase transition for a large enough $\mathbf{A}(\mathbf{r})^2$ term \cite{Birula79,Nataf10b}. The standard understanding of this phenomenon is that, when a single photonic mode is coupled to the dipolar transition, the $\mathbf{A}(\mathbf{r})^2$ term shifts it toward higher energies, making it impossible to reach the critical resonant coupling. In the present work, performing a careful multimode analysis that allowed us to study the process in real space, we showed that such a shift is only an aspect of a more general phenomenon, hampering not only the possibility to observe superradiant phase transitions, but also any ultrafast process based on energy exchange between light and matter degrees of freedom. The light-matter decoupling, and the consequent breakdown of the Purcell effect, cannot generally be attributed to a simple shift of the photonic modes toward higher frequencies, as light and matter decouple even in presence of cavity modes quasi-resonant with the matter one ({\it e.g.}, in Fig. \ref{MainFig} (a), for $\Omega=2\omega_0$, the matter mode and the first photonic mode detuning is much smaller than $\Omega$). 

It is important to notice that, while the $\mathbf{A}(\mathbf{r})^2$ is normally always present in the light-matter coupling Hamiltonian, as it is due to the fundamental form of the interaction between electrons and photons, it can be artificially suppressed in nonequilibrium systems \cite{Baumann10,Bastidas12,Bhaseen12}. 
Moreover it has recently been proposed that in some systems, like superconducting qubits \cite{Nataf10b} and graphene \cite{Hagenmuller12}, the $\mathbf{A}(\mathbf{r})^2$ term could also be suppressed, even if there is still no consensus in the community on the accuracy of these predictions \cite{Chirolli13,Viehmann11,Ciuti12}.
As the  $\mathbf{A}(\mathbf{r})^2$ is modified, the light-matter decoupling effect will probably be affected.
It will thus be most interesting to investigate to which extent the theory developed in the present work remains valid for these systems.

\section{Analytic formula for the spectrum of the Light-Matter Hamiltonian}
\label{Diagonalisation}

In this Appendix we calculate the spectrum of a photonic cavity coupled to a bosonic matter degree of freedom.
To this aim we introduce the bosonic annihilation operator $a_n$ for the $n^{\text{th}}$ mode of the cavity, with frequency $\omega_n$. Such a mode will couple to the matter mode $b$, of  frequency $\omega_0$,
through the coupling constant $\Omega_n$.
The Hamiltonian of the system thus reads
\begin{eqnarray}
\label{Hsm}
H_{\text{LM}}&=&\hbar\omega_0 b^{\dagger} b + \sum_n \hbar\omega_n a^{\dagger}_na_n + \sum_{n} \hbar\Omega_n
(b^{\dagger}+b)(a^{\dagger}_n+a_n)\nonumber \\&&+\sum_{n,m} \frac{\hbar\Omega_n\Omega_m}{\omega_0}(a^{\dagger}_n+a_n)(a^{\dagger}_m+a_m).
\end{eqnarray}
It is convenient to introduce the generalised position and momentum operators
\begin{eqnarray}
\label{XP}
X&=&\frac{b^{\dagger}+b}{\sqrt{2}},\quad\quad
P=i\frac{b^{\dagger}-b}{\sqrt{2}}, \quad \\ 
Y_n&=&\frac{a^{\dagger}_n+a_n}{\sqrt{2}}, \quad
Q_n=i\frac{a^{\dagger}_n-a_n}{\sqrt{2}}. \nonumber
\end{eqnarray}
The Heisenberg equations for such operators read
\begin{eqnarray}
\label{dotX}
\dot{X}&=&\omega_0 P, \quad \dot{P}=-\omega_0 X -\sum_n 2\Omega_n Y_n, \\
\dot{Y}_n&=&\omega_n Q_n,\quad \dot{Q}_n=-\omega_n Y_n-\sum_m \frac{4\Omega_n\Omega_m}{\omega_0}Y_m-2\Omega_n X\nonumber. 
\end{eqnarray}
Deriving a second time, passing in Fourier space, and introducing the two quantities
\begin{eqnarray}
\label{Zf}
\tilde{Z}(\omega)&=&\sum_n 2\Omega_n \omega_n \tilde{Q}_n(\omega),\\
f(\omega)&=&\sum_n \frac{4\Omega_n^2}{\omega_0\omega_n (1-\frac{\omega^2}{\omega_n^2})},\nonumber
\end{eqnarray}
we can extract from the system in \Eq{dotX} two closed algebraic equations
\begin{eqnarray}
\label{PZ}
\tilde{P}(\omega)(\omega_0^2-\omega^2)&=&-\tilde{Z}(\omega),\\
\tilde{Z}(\omega)\lbrack 1+f(\omega)\rbrack&=&-\omega_0^2 f(\omega) \tilde{P}(\omega)\nonumber.
\end{eqnarray}
The eigenfrequencies of the Hamiltonian in \Eq{Hsm} are thus given by the solutions of the equation
\begin{eqnarray}
\label{Eqsm}
\omega_0^2-\omega^2=\omega^2 f(\omega).
\end{eqnarray}
We will now specialise the formula in \Eq{Eqsm} to the case of a perfect metallic cavity of length $L_{\text{C}}$, with a two dimensional wall of dipoles at distance $L_{\text{W}}=lL_{\text{C}}$ from one of the mirrors. 
As the photonic modes are in this case sinusoidal and equispaced, we have $\omega_n=n\omega_{\text{c}}$ and $\Omega_n=\frac{\Omega\sin(\pi l n)}{\sqrt{n}}$. From the second line of \Eq{Zf} we get
\begin{eqnarray}
\label{f}
f(\omega)&=&\frac{4\Omega^2}{\omega_{\text{c}}\omega_0} \sum_n \frac{\sin(n\pi l)^2}{(n^2-\frac{\omega^2}{\omega_{\text{c}}^2})}\\&=&
\frac{2\pi \Omega^2}{\omega_0\omega} \frac{\sin(\pi l \frac{\omega}{\omega_{\text{c}}})\sin(\pi (1-l) \frac{\omega}{\omega_{\text{c}}})}{\sin(\pi \frac{\omega}{\omega_{\text{c}}})},\nonumber
\end{eqnarray}
where the summation in the last step can be performed by calculating the residues of the function
\begin{eqnarray}
g(z)&=&\frac{\sin(\pi lz)\sin(\pi  (1-l)z)}{\sin(\pi z)(z^2-\frac{\omega^2}{\omega_{\text{c}}^2})}.
\end{eqnarray}
The spectrum of the Hamiltonian in \Eq{Hsm} is thus given by the solutions of the following transcendental equation
\begin{eqnarray}
\label{Eqfsm}
\omega_0^2-\omega^2=2\pi \Omega^2 \frac{\omega}{\omega_0} \frac{\sin(\pi l \frac{\omega}{\omega_{\text{c}}})\sin(\pi (1-l) \frac{\omega}{\omega_{\text{c}}})}{\sin(\pi \frac{\omega}{\omega_{\text{c}}})}.
\end{eqnarray}

\section{Hopfield-Bogoliubov diagonalization of the light-matter Hamiltonian}
\label{HBdiag}
The light-matter Hamiltonian in \Eq{Hfull}, and its more general form in \Eq{Hsm}, are quadratic and bosonic, it is thus possible to diagonalise them using an Hopfield-Bogoliubov procedure \cite{Hopfield58}.
The diagonalised Hamiltonian will describe a set of free bosonic modes: mixed light and matter excitations, usually called polaritons. Considering any quantity with an index as a line vector, any quantity with two indexes as a matrix, and introducing the vector
\begin{eqnarray}
\mathbf{v}&=&\lbrack b,\, a_n, \, b^{\dagger},\, a_n^{\dagger} \rbrack,
\end{eqnarray}
we can rewrite the light-matter Hamiltonian as
\begin{eqnarray}
\label{HOB}
H_{\text{LM}}&=&\frac{\hbar}{2}
\mathbf{v}^{\dagger}\eta M\mathbf{v},
\end{eqnarray}
where 
\begin{eqnarray}
\eta&=&\text{diag}\lbrack 1,\,\dots\,,1,-1,\,\dots\,,-1\rbrack,
\end{eqnarray}
is a diagonal metric matrix, and the Hopfield-Bogoliubov matrix $M$ is given by the expression
\begin{eqnarray}
\label{M}
\hspace{-1cm}
M&=&
\left[
\begin{array}{cccc}
\omega_0, & \Omega_n, & 0 & \Omega_n\\
\Omega_n^{\dagger},& \omega_n \delta_{nm}+\frac{2\Omega_n\Omega_m}{\omega_0},& \Omega_n^{\dagger},&\frac{2\Omega_n\Omega_m}{\omega_0}\\
0,& -\Omega_n& -\omega_0, & -\Omega_n\\
-\Omega_n^{\dagger}&-\frac{2\Omega_n\Omega_m}{\omega_0},&-\Omega_n^{\dagger},&- \omega_n \delta_{nm}-\frac{2\Omega_n\Omega_m}{\omega_0}
\end{array}
\right].\nonumber \\
\end{eqnarray}
The matrix $M$ can be diagonalised, its eigenvalues yielding the polaritonic frequencies $\omega_{\mu}$, and the relative eigenvectors the decomposition of the polaritonic operators over the operators of the uncoupled light and matter modes
\begin{eqnarray}
\label{pSup}
p_{\mu}&=&U_{\mu 0}b+\sum_n U_{\mu n}a+ V_{\mu 0}b^{\dagger}+\sum_n V_{\mu n}a^{\dagger}.
\end{eqnarray}
We finally obtain
\begin{eqnarray}
\label{Hsup}
H_{\text{LM}}&=&\sum_{\mu} \hbar \omega_{\mu} p^{\dagger}_{\mu}p_{\mu}.
\end{eqnarray}
Notice that, in order for the polaritonic operators to satisfy bosonic commutation relations
\begin{eqnarray}
\label{commp}
\lbrack p_{\mu},\, p_{\mu'}^{\dagger}\rbrack&=&\delta_{\mu\mu'},
\end{eqnarray}
the coefficients in \Eq{pSup}, that without any loss of generality we can suppose to be real, have to satisfy the $\eta$-normalisation relation
\begin{eqnarray}
\label{norm}
U^2_{\mu 0}+\sum_n U_{\mu n}^2- V_{\mu 0}^2-\sum_n V_{\mu n}^2&=&1.
\end{eqnarray}
Moreover the the $\eta$-orthonormality of the polaritonic modes implies that we can easily invert \Eq{pSup} as
\begin{eqnarray}
\label{pSupInv}
a_{n}&=&\sum_{\mu} U_{\mu n}p_{\mu}-V_{\mu n}p^{\dagger}_{\mu},\\
b&=&\sum_{\mu} U_{\mu 0}p_{\mu}-V_{\mu 0}p^{\dagger}_{\mu}.\nonumber 
\end{eqnarray}

\section{Classical Electromagnetic Theory}
\label{ClassicalTheory}
Both the spectrum of the light-matter system and the profile of the electromagnetic modes inside the cavity can be derived from classical electromagnetic theory using a transfer matrix approach \cite{Kavokin}.
The transfer matrix for the propagation of an electromagnetic wave with wavevector $k=\omega/c$ in a medium of length $L$ is
\begin{eqnarray}
T_{P}(\omega,L)&=&
\left[
\begin{array}{cc}
 e^{i\omega L/c}&  0  \\
0 & e^{-i\omega L/c}      
\end{array}
\right],
\end{eqnarray}
and the one for the dipolar wall is
\begin{eqnarray}
T_{\text{W}}(\omega)&=&
\frac{1}{t(\omega)}\left[
\begin{array}{cc}
 t(\omega)^2-r(\omega)^2&  r(\omega)  \\
-r(\omega) & 1      
\end{array}
\right],
\end{eqnarray}
where $r(\omega)$ and $t(\omega)$ are the reflection and transmission coefficients of the wall.
Their explicit expressions are \cite{Kavokin}
\begin{eqnarray}
\label{rsm}
r(\omega)&=&\frac{i\pi\Omega^2}{\frac{\omega_{\text{0}}}{\omega}(\omega_0^2-\omega^2)-i\pi\Omega^2},\nonumber \\
t(\omega)&=&1+r(\omega).
\end{eqnarray}
The total transfer matrix for the system will thus be
\begin{eqnarray}
T(\omega)&=&T_P(\omega,L_{\text{W}})T_{\text{W}}(\omega)T_P(\omega,L_{\text{C}}-L_{\text{W}}),
\end{eqnarray}
and the spectrum of the system can be found imposing metallic boundary conditions, that is
\begin{eqnarray}
T(\omega)
\left[
\begin{array}{c}
    -1    \\
      1  
\end{array}
\right]&=&A
\left[
\begin{array}{c}
    1    \\
      -1  
\end{array}
\right].
\label{Classical2}
\end{eqnarray}
Eliminating the coefficient $A$ from \Eq{Classical2} we obtain the transcendental equation 
\begin{eqnarray}
\frac{T^{21}(\omega)-T^{22}(\omega)}{T^{12}(\omega)-T^{11}(\omega)}=1,
\end{eqnarray}
that can be written explicitly as
\begin{eqnarray}
\label{inter}
\omega_0^2-\omega^2=\frac{2\omega \pi \Omega^2}{\omega_0}\frac{\sin(\omega L_{\text{W}}/c)\sin (\omega(L_{\text{C}}-L_{\text{W}})/c)}{\sin(\omega L_{\text{C}}/c)}.\quad
\end{eqnarray}
Upon the substitution $\omega_c=\frac{c\pi}{L_{\text{C}}}$ for the frequency of the fundamental cavity mode, we obtain exactly the formula  found with a quantum approach in 
Appendix \ref{Diagonalisation}.
Once a solution $\bar{\omega}$ has been found from \Eq{inter}, its electromagnetic mode profile can be determined propagating the 
metallic boundary condition as
\begin{eqnarray}
\label{Esm}
\mathbf{E}(z)&=&\mathbf{e}_{\parallel}(f^+ + f^-),
\end{eqnarray}
where
\begin{eqnarray}
\left[
\begin{array}{c}
    f^+    \\
     f^-  
\end{array}
\right]&=&T(\bar{\omega},z)
\left[
\begin{array}{c}
    1    \\
      -1  
\end{array}
\right],
\label{Classical}
\end{eqnarray}
and
\begin{eqnarray}
T(\bar{\omega},z)&=&T_{P}(\bar{\omega},z),\quad 0<z<L_{\text{W}},\\
T(\bar{\omega},z)&=&T_{P}(\bar{\omega},L_{\text{W}})T_{\text{W}}T_{P}(\bar{\omega},z-L_{\text{W}}),\quad L_{\text{W}}<z<L_{\text{C}}.\nonumber
\end{eqnarray}
Notice that, even if this classical approach reproduces exactly the profile of the modes plotted in the main body of the paper, it cannot give their relative normalisations, nor describe the fraction of the energy stored in the matter part of the excitations.

\section{Input-Output Theory}
\label{In-out}
\begin{figure*}[t]
\begin{center}
\includegraphics[width=18.0cm]{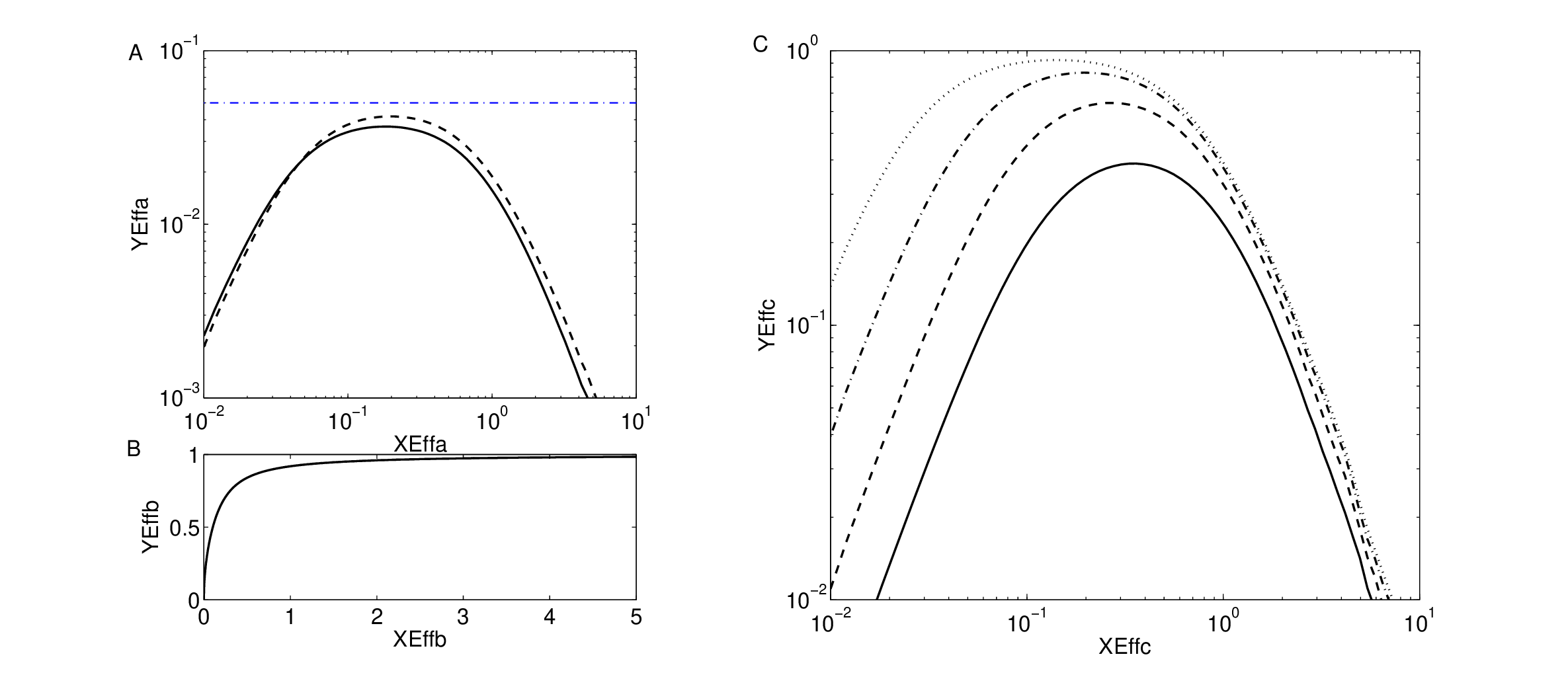}
\caption{\label{Cont} Panel (a): spontaneous emission rate as a function of the normalized light-matter coupling. The black dashed line is obtained using frequency independent parameters $\kappa_{\text{ph}}(\omega)=\kappa_{\text{el}}(\omega)=\kappa$. The black solid line is instead obtained using frequency dependent ones $\kappa_{\text{ph}}(\omega)=\kappa_{\text{el}}(\omega)=\kappa(\omega)$. The blue dash-dotted line is the saturation value $\gamma_{\text{sat}}$ obtained for frequency independent parameters. Panel (b): frequency dependence of the function $\kappa(\omega)$ used in panel (a). Panel (c): spontaneous emission rate, normalized to the saturation value $\gamma_{\text{sat}}$, as a function of the normalized light-matter coupling, for different values of the loss coefficients $\Gamma_{\text{ph}}=\Gamma_{\text{el}}=0.2$ (solid line), $0.1$ (dashed line), $0.05$ (dashed-dotted line), and $0.025$ (dotted line). Unless otherwise stated, the same parameters of Fig. 4 have been used.}
\end{center}
\end{figure*}

In order to calculate the extra-cavity observables, we couple the system to its environment, described with two harmonic baths. The first models the extra-cavity photonic modes $\alpha_{\omega}$, that couple to the intra-cavity $a_n$ modes through the finite reflectivity of the metallic mirrors, and the second the electronic excitations $\beta_{\omega}$, capable to excite the $b$ matter mode. We can describe them through the Hamiltonians
\begin{eqnarray}
\label{BPHSup}
H_{\text{ph}}&=&\hbar\int_0^{\infty} d\omega\, \omega\, \alpha_{\omega}^{\dagger}\alpha_{\omega} +\sum_{n}\kappa_{\text{ph}}(\omega) \lbrack a_{n}^{\dagger}\alpha_{\omega} +\alpha_{\omega}^{\dagger}a_{n}\rbrack, \nonumber \\
H_{\text{el}}&=&\hbar\int_0^{\infty}  d\omega\, \omega\, \beta_{\omega}^{\dagger}\beta_{\omega} +\kappa_{\text{el}}(\omega) \lbrack b_{n}^{\dagger}\beta_{\omega} +\beta_{\omega}^{\dagger}b_{n}\rbrack,
\end{eqnarray}
with
\begin{eqnarray}
\lbrack \alpha_{\omega},\,\alpha_{\omega'}^{\dagger}\rbrack&=&\lbrack \beta_{\omega},\,\beta_{\omega'}^{\dagger}\rbrack=\delta(\omega-\omega').
\end{eqnarray}
Using \Eq{pSupInv} we can rewrite these Hamiltonians in terms of the polaritonic operators as
\begin{eqnarray}
\label{BPHPolSup}
H_{\text{ph}}&=&\hbar\int_0^{\infty} d\omega\, \omega\, \alpha_{\omega}^{\dagger}\alpha_{\omega} +\sum_{n,\mu}\kappa_{\text{ph}}(\omega)U_{\mu n} \lbrack p_{\mu}^{\dagger}\alpha_{\omega} +\alpha_{\omega}^{\dagger}p_{\mu}\rbrack \nonumber \\&&
-\sum_{n,\mu}\kappa_{\text{ph}}(\omega)V_{\mu n} \lbrack p_{\mu}\alpha_{\omega} +\alpha_{\omega}^{\dagger}p_{\mu}^{\dagger}\rbrack, \nonumber \\
H_{\text{el}}&=&\hbar\int_0^{\infty}  d\omega\, \omega\, \beta_{\omega}^{\dagger}\beta_{\omega} +\sum_{\mu}\kappa_{\text{el}}(\omega)U_{\mu 0} \lbrack p_{\mu}^{\dagger}\beta_{\omega} +\beta_{\omega}^{\dagger}p_{\mu}\rbrack \nonumber \\&&
-\sum_{\mu}\kappa_{\text{el}}(\omega)V_{\mu 0} \lbrack p_{\mu}\beta_{\omega} +\beta_{\omega}^{\dagger}p_{\mu}^{\dagger}\rbrack.
\end{eqnarray}
As shown in Ref. \cite{Bamba13}, the antiresonant terms in \Eq{BPHPolSup}, consisting of two annihilation or two creation operators, are not present in a rigorous microscopic approach, as the negative and positive frequency components of the electromagnetic field do not mix (and the same can be safely assumed for the electronic excitations). We have thus to perform a rotating wave approximation (RWA) on the system-environment coupling in order to eliminate the antiresonant terms. 
In order to do that, we cannot simply neglect the $V_{0\mu}$ and $V_{n\mu}$ terms, because
the polaritonic modes need to be $\eta$-normalised, as from \Eq{norm} and, due to the minus signs in  \Eq{norm},  this would lead to unnormalised (and thus non-bosonic) polaritonic operators.   
We thus need to renormalise the polaritonic modes after having performed the RWA, obtaining
\begin{eqnarray}
\label{BPHNorSup}
H_{\text{ph}}&=&\hbar\int_0^{\infty} d\omega\, \omega\, \alpha_{\omega}^{\dagger}\alpha_{\omega} +\sum_{n,\mu}\bar{\kappa}^{\mu}_{\text{ph}}(\omega) \lbrack p_{\mu}^{\dagger}\alpha_{\omega} +\alpha_{\omega}^{\dagger}p_{\mu}\rbrack, \nonumber \\
H_{\text{el}}&=&\hbar\int_0^{\infty}  d\omega\, \omega\, \beta_{\omega}^{\dagger}\beta_{\omega} +\sum_{\mu}\bar{\kappa}^{\mu}_{\text{el}}(\omega) \lbrack p_{\mu}^{\dagger}\beta_{\omega} +\beta_{\omega}^{\dagger}p_{\mu}\rbrack,\quad\quad\,\,
\end{eqnarray}
where
\begin{eqnarray}
\label{KT}
\bar{\kappa}_{\text{ph}}^{\mu}(\omega)&=&\sum_{n} \frac{U_{\mu n}}{\sqrt{U_{\mu 0}^2+\sum_n U_{\mu n}^2}}\kappa_{\text{ph}}(\omega),\\
\bar{\kappa}_{\text{el}}^{\mu}(\omega)&=& \frac{U_{\mu 0}}{\sqrt{U_{\mu 0}^2+\sum_n U_{\mu n}^2}}\kappa_{\text{el}}(\omega).
\end{eqnarray}
Such an approach has been shown in Ref. \cite{DeLiberato13b} to give the same results of a microscopic approach based on Maxwell boundary conditions.
We can now write the dynamical equations for the extra-cavity electromagnetic field, $\alpha_{\omega}$, in the form 
\begin{equation}
\dot{\alpha}_{\omega}=-i\omega\alpha_{\omega}-i\sum_{\mu}\bar{\kappa}_{\text{ph}}^{\mu}(\omega)p_{\mu},
\end{equation}
whose solution can be formally written as
\begin{eqnarray}
\label{Bio}
\alpha_{\omega}(t)&=&e^{-i\omega(t-t_0)}\alpha_{\omega}(t_0)\\\nonumber&&
-i\sum_{\mu}\bar{\kappa}_{\text{ph}}^{\mu}(\omega)\int_{t_0}^{t}dt' e^{-i\omega(t-t')}  p_{\mu}(t'),
\end{eqnarray}
$t_0$ being an arbitrary initial time. 
This formula can be inserted into the evolution equation for the polaritonic operators
\begin{eqnarray}
\label{dinamicaa}
\dot{p}_{\mu}&=&-i\omega_{\mu}p_{\mu}\\&& \nonumber -i \sum_{\mu}\int_0^{\infty} d\omega\, \lbrack \bar{\kappa}_{\text{ph}}^{\mu}(\omega) \alpha(\omega)+
\bar{\kappa}_{\text{el}}^{\mu}(\omega) \beta(\omega)\rbrack.
\end{eqnarray} 
Doing the same for the electronic $\beta_{\omega}$ operators, using the standard definitions
\begin{eqnarray}
\label{BdefSup}
\alpha_{\omega}^{\text{in}}&\equiv& \lim_{t \rightarrow -\infty}\alpha_{\omega}(t)e^{-i\omega t},\quad \beta_{\omega}^{\text{in}}\equiv \lim_{t \rightarrow -\infty}\beta_{\omega}(t)e^{-i\omega t}, \nonumber \\
\alpha_{\omega}^{\text{out}}&\equiv& \lim_{t \rightarrow \infty}\alpha_{\omega}(t)e^{-i\omega t},\quad \beta_{\omega}^{\text{out}}\equiv \lim_{t \rightarrow \infty}\beta_{\omega}(t)e^{-i\omega t},\quad\quad\quad 
\end{eqnarray}
for the input and output fields, 
and introducing the decay rates and the Langevin forces 
\begin{eqnarray}
\label{Gammas}
\Gamma_{\text{ph}}^{\mu\mu'}(t)&=&\Theta(t) \int_0^{\infty} d\omega \,
\bar{\kappa}_{\text{ph}}^{\mu}(\omega) \bar{\kappa}_{\text{ph}}^{\mu'}(\omega)e^{-i\omega t},\\ \nonumber
\Gamma_{\text{el}}^{\mu\mu'}(t)&=&\Theta(t) \int_0^{\infty} d\omega \,
\bar{\kappa}_{\text{el}}^{\mu}(\omega) \bar{\kappa}_{\text{el}}^{\mu'}(\omega) e^{-i\omega t},\\ \nonumber
F^{\mu}_{\text{ph}}(t)&=&
\int_0^{\infty} d\omega\, \bar{\kappa}_{\text{ph}}^{\mu}(\omega) \alpha_{\omega}^{\text{in}}e^{-i\omega t},\\\nonumber
F^{\mu}_{\text{el}}(t)&=&
\int_0^{\infty} d\omega\, \bar{\kappa}_{\text{el}}^{\mu}(\omega) \beta_{\omega}^{\text{in}}e^{-i\omega t},
\end{eqnarray}
with $\Theta(t)$ the Heaviside function, 
one can cast \Eq{dinamicaa} in the form of a quantum Langevin equation
\begin{eqnarray}
\label{langevinta}
\dot{p}_{\mu}&=&-i\omega_{\mu}p_{\mu}-\int_{-\infty}^{\infty} dt' \lbrack\Gamma_{\text{ph}}^{\mu\mu'}(t-t')+\Gamma_{\text{el}}^{\mu\mu'}(t-t')\rbrack p_{\mu'}(t')\nonumber\\&&
-iF_{\text{ph}}^{\mu}(t)-iF_{\text{el}}^{\mu}(t).
\end{eqnarray}
Passing in Fourier space, it is possible to formally solve the system in \Eq{langevinta} as
\begin{eqnarray}
\label{pG}
\tilde{p}_{\mu}(\omega)&=&-iG_{\mu\mu'}(\omega)\lbrack \tilde{F}^{\mu'}_{\text{ph}}(\omega)+\tilde{F}^{\mu'}_{\text{el}}(\omega)\rbrack.
\end{eqnarray}
Inserting \Eq{pG} into \Eq{Bio}, we finally obtain a set of linear relations between the extra-cavity input and output fields
\begin{eqnarray}
\left[
\begin{array}{c}
 \alpha^{\text{out}}_{\omega}     \\
 \beta^{\text{out}}_{\omega} 
\end{array}
\right]
&=&
\left[
\begin{array}{cc}
 \mathcal{U}^{11}(\omega) &   \mathcal{U}^{12}(\omega)  \\
 \mathcal{U}^{21}(\omega)& \mathcal{U}^{22}(\omega)      
\end{array}
\right]
\left[
\begin{array}{c}
 \alpha^{\text{in}}_{\omega}    \\
 \beta^{\text{in}}_{\omega} 
\end{array}
\right],
\end{eqnarray}
with
\begin{eqnarray}
\mathcal{U}^{11}(\omega)&=&1-2\pi \sum_{\mu,\mu'} 
\bar{\kappa}_{\text{ph}}^{\mu}(\omega) \bar{\kappa}_{\text{ph}}^{\mu'}(\omega) 
G_{\mu\mu'}(\omega),\nonumber \\
\mathcal{U}^{12}(\omega)&=&-2\pi\sum_{\mu,\mu'} 
\bar{\kappa}_{\text{ph}}^{\mu}(\omega) \bar{\kappa}_{\text{el}}^{\mu'}(\omega) 
G_{\mu\mu'}(\omega),\nonumber \\
\mathcal{U}^{22}(\omega)&=&1-2\pi\sum_{\mu,\mu'}
\bar{\kappa}_{\text{el}}^{\mu}(\omega) \bar{\kappa}_{\text{el}}^{\mu'}(\omega) 
 G_{\mu\mu'}(\omega), \nonumber \\
 \mathcal{U}^{21}(\omega)&=&-2\pi\sum_{\mu,\mu'} 
\bar{\kappa}_{\text{el}}^{\mu}(\omega) \bar{\kappa}_{\text{ph}}^{\mu'}(\omega) 
G_{\mu\mu'}(\omega).
\end{eqnarray}
The scattering properties of the coupled light-matter system are thus determined once the coupling parameters to the extracavity fields are fixed. In the simulations we used frequency-independent parameters, but we verified that the  results are not qualitatively modified using continuous parameters such that $\kappa_{\text{ph}}(0)={\kappa}_{\text{el}}(0)=0$ (a direct comparison can be found in the panels (a) and (b) of Fig. \ref{Cont}). Moreover, we neglected the Lamb shift, due to the imaginary part of the Fourier transform of the decay rates in \Eq{Gammas}, as it is expected to be completely negligible, and its exact form strongly depends on microscopic and system-dependent details. 
\psfrag{XEffa}[B][t][1]{${\Omega/\omega_0}$}
\psfrag{XEffb}[T][c][1]{${\omega/\omega_0}$}
\psfrag{YEffa}[T][t][1]{${\gamma/\omega_0}$}
\psfrag{YEffb}[B][t][1]{${\kappa(\omega)/\kappa} $ }
\psfrag{XEffc}[T][t][1]{${\Omega/\omega_0}$}
\psfrag{YEffc}[B][t][1]{${\gamma/\gamma_{\text{sat}}}$}
\psfrag{A}[T][T][1]{\hspace{0cm}$\mathbf{(a)}$}
\psfrag{B}[T][B][1]{\hspace{0cm}$\mathbf{(b)}$}
\psfrag{C}[T][B][1]{\hspace{0cm}$\mathbf{(c)}$}
In order to better explore the effect of dissipation upon the spontaneous emission rate, in Fig. \ref{Cont}(c) we plot the spontaneous emission rate $\gamma$, normalized to its saturation value $\gamma_{\text{sat}}$, for various values of the loss coefficients.  We see that, while the qualitative behaviour of the system remains the same, confirming that the decoupling effect is solid against dissipation, the maximal value of the emission rate decreases from the theoretical value $\gamma_{\text{sat}}$ as the dissipation increases. This can be understood noticing that the quadratic, Purcell-like spontaneous emission rate valid in the weak and strong coupling regimes
\begin{eqnarray}
\label{LumPAP}
\gamma_{\text{Pur}}&\simeq&\frac{2\Omega^2}{\Gamma_{\text{ph}}+\Gamma_{\text{el}}},
\end{eqnarray}
crosses the saturation value 
\begin{eqnarray}
\label{LumSAP}
\gamma_{\text{sat}}&\simeq&\frac{2\Gamma_{\text{ph}}\Gamma_{\text{el}}}{\Gamma_{\text{ph}}+\Gamma_{\text{el}}}
\end{eqnarray}
for 
\begin{eqnarray}
\label{ConsAP}
\Omega=\sqrt{\Gamma_{\text{ph}}\Gamma_{\text{el}}}.
\end{eqnarray}
In order to saturate at the theoretical maximum value in \Eq{LumSAP}, the crossing needs to be in a region in which \Eq{LumPAP} is still valid, that is roughly for a normalized coupling smaller than $0.1$. If this is not the case, due to the presence of strong dissipation, the emission rate will saturate at a value lower than $\gamma_{\text{sat}}$.


\begin{thebibliography}{}
\bibitem{Purcell46} E. M. Purcell, Phys. Rev. {\bf 69}, 681 (1946).
\bibitem{Haroche} S. Haroche and J.-M. Raimond, {\it Exploring the Quantum: Atoms, Cavities, and Photons}, Oxford University Press (2006).
\bibitem{Kavokin} A.V. Kavokin, J. J. Baumberg, G. Malpuech, and F. P. Laussy, {\it Microcavities}, Oxford University Press (2011).
\bibitem{Devoret07} M. Devoret, S. Girvin, and R. Schoelkopf, Ann. Phys. (Leipzig) {\bf 16}, 767 (2007).
\bibitem{Ciuti05} C. Ciuti, G. Bastard and I. Carusotto, Phys. Rev. B {\bf 72}, 115303 (2005).
\bibitem{Ciuti06} C. Ciuti, I. Carusotto, Phys. Rev. A {\bf 74}, 033811 (2006).
\bibitem{Dini03} D. Dini, R. Kohler, A. Tredicucci, G. Biasiol and L. Sorba, Phys. Rev. Lett. {\bf 90}, 116401 (2003).
\bibitem{Auer12} A. Auer and G. Burkard, Phys. Rev. B {\bf 85}, 235140 (2012).
\bibitem{DeLiberato07} S. De Liberato, C. Ciuti and I. Carusotto, Phys. Rev. Lett. {\bf 98}, 103602 (2007).
\bibitem{Lambert04} N. Lambert, C. Emary and T. Brandes, Phys. Rev. Lett. {\bf 92}, 073602 (2004). 
\bibitem{Nataf10b} P. Nataf and C. Ciuti, Nat. Comm. {\bf 1}, 72 (2010).
\bibitem{DeLiberato13} S. De Liberato and C. Ciuti, Phys. Rev. Lett. {\bf 110}, 133603 (2013).
\bibitem{DeLiberato08}  S. De Liberato and C. Ciuti, Phys. Rev. B {\bf 77}, 155321 (2008). 
\bibitem{DeLiberato09a} S. De Liberato and C. Ciuti, Phys. Rev. B {\bf 79}, 075317 (2009).
\bibitem{Niemczyk10} T. Niemczyk, F. Deppe, H. Huebl, E. P. Menzel, F. Hocke, M. J. Schwarz, J. J. Garcia-Ripoll, D. Zueco, T. Hummer, E. Solano, A. Marx, and R. Gross, Nat. Phys. {\bf 6}, 772 (2010).
\bibitem{Muravev11} V. M. Muravev, I. V. Andreev, I. V. Kukushkin, S. Schmult, and W. Dietsche, Phys. Rev. B {\bf 83}, 075309 (2011).
\bibitem{Anappara09} A. A. Anappara, S. De Liberato, A. Tredicucci, C. Ciuti, G. Biasiol, L. Sorba and F. Beltram, Phys. Rev. B {\bf 79}, 201303 (2009).
\bibitem{Todorov10} Y. Todorov, A. M. Andrews, R. Colombelli, S. De Liberato, C. Ciuti, P. Klang, G. Strasser and C. Sirtori, Phys. Rev. Lett. {\bf 105}, 196402 (2010).
\bibitem{Geiser12} M. Geiser, F. Castellano, G. Scalari, M. Beck, L. Nevou, and J. Faist, Phys. Rev. Lett. {\bf 108}, 106402 (2012).
\bibitem{Schwartz11} T. Schwartz, J. A. Hutchison, C. Genet and T. W. Ebbesen, Phys. Rev. Lett. {\bf 106}, 196405 (2011).
\bibitem{Scalari12} G. Scalari, C. Maissen, D. Turcinkova, D. Hagenm\"uller, S. De Liberato, C. Ciuti, C. Reichl, D. Schuh, W. Wegscheider, M. Beck, and J. Faist, Science {\bf 335}, 1323 (2012).
\bibitem{Jouy10} P. Jouy, A. Vasanelli, Y. Todorov, L. Sapienza, R. Colombelli, U. Gennser, and C. Sirtori, Phys. Rev. B {\bf 82}, 045322 (2010).
\bibitem{Bamba13} M. Bamba and T. Ogawa, Phys. Rev. A {\bf 88}, 013814 (2013).
\bibitem{Casanova10} J. Casanova, G. Romero, I. Lizuain, J. J. Garc\'ia-Ripoll, and E. Solano, Phys. Rev. Lett. {\bf 105}, 263603 (2010).
\bibitem{Hagenmuller10} D. Hagenm\"uller, S. De Liberato and C. Ciuti, Phys. Rev. B {\bf 81}, 235303 (2010).
\bibitem{Beaudoin11} F. Beaudoin, J. M. Gambetta and A. Blais, Phys. Rev. A {\bf 84}, 043832 (2011).
\bibitem{Birula79} I. Bialynicki-Birula and K. Rz\c a\.znewski, Phys. Rev. A {\bf 19}, 301 (1979).
\bibitem{DeLiberato13b} S. De Liberato, Phys. Rev. A {\bf 89}, 017801 (2014).
\bibitem{Anappara05} A. A. Anappara, A. Tredicucci, G. Biasiol, and L. Sorba, Appl. Phys. Lett. {\bf 87}, 051105 (2005). 
\bibitem{Gunter09} G. G\"unter, A. A. Anappara, J. Hees, A. Sell, G. Biasiol, L. Sorba, S. De Liberato, C. Ciuti, A. Tredicucci, A. Leitenstorfer and R. Huber, Nature {\bf 458}, 178 (2009). 
\bibitem{Porer12} M. Porer, J.-M Menard, A. Leitenstorfer, R. Huber, R. Degl'Innocenti, S. Zanotto, G. Biasiol, L. Sorba, and A. Tredicucci, Phys. Rev. B {\bf 85}, 081302 (2012).
\bibitem{Hopfield58} J. J. Hopfield, Phys. Rev. {\bf 112}, 1555 (1958).
\bibitem{Bastidas12} V. M. Bastidas, C. Emary, B. Regler, and T. Brandes, Phys. Rev. Lett. {\bf 108}, 043003 (2012).
\bibitem{Bhaseen12} M. J. Bhaseen, J. Mayoh, B. D. Simons, and J. Keeling, Phys. Rev. A {\bf 85}, 013817 (2012).
\bibitem{Baumann10} K. Baumann, C. Guerlin,  F. Brennecke, and T. Esslinger, Nature {\bf 464}, 1301 (2010).
\bibitem{Hagenmuller12} D. Hagenm\"uller and C. Ciuti, Phys. Rev. Lett. {\bf 109}, 267403 (2012).
\bibitem{Chirolli13} L. Chirolli, M. Polini, V. Giovannetti, and A. H. MacDonald, Phys. Rev. Lett. {\bf 109}, 267404 (2012).
\bibitem{Viehmann11} O. Viehmann, J. von Delft, and F. Marquardt, Phys. Rev. Lett. {\bf 107}, 113602 (2011).
\bibitem{Ciuti12} C. Ciuti and P. Nataf, Phys. Rev. Lett. {\bf 109}, 179301 (2012).

\end{thebibliography}
\end{document}